\documentclass[12pt]{article}

\newcommand{\VersionInformation}{}  
\InputIfFileExists{Debug}{}{}

\usepackage[title,titletoc]{appendix}
\usepackage{enumerate}
\usepackage[utf8]{inputenc}
\usepackage{amsmath}
\usepackage{amsthm}
\usepackage{amsfonts}          
\usepackage{amssymb}           
\usepackage[mathscr]{eucal}
\usepackage{graphicx}
\usepackage{color}
\usepackage{hypbmsec}
\usepackage{multirow}
\usepackage[vcentermath]{youngtab}
\usepackage[isu,bf]{caption}
\newlength{\xtrawidth}
\newlength{\xtraheight}
\ifx\NoBackreferences\EMPTYMACRO
  \usepackage[backref,linktocpage,bookmarks,plainpages=false]{hyperref}
\else
  \usepackage[linktocpage,bookmarks]{hyperref}
\fi

\usepackage[compress]{cite}

\ifx\DEBUG\EMPTYMACRO
\else
  \usepackage{showlabels}
\fi

\def\clap#1{\hbox to 0pt{\hss#1\hss}}

\makeatletter
  \def\adots{\mathinner{\mkern2mu\raise\p@\hbox{.}
      \mkern2mu\raise4\p@\hbox{.}\mkern1mu
      \raise7\p@\vbox{\kern7\p@\hbox{.}}\mkern1mu}}
\makeatother

\newcommand{\R}{\ensuremath{{\mathbb{R}}}}
\newcommand{\C}{\ensuremath{{\mathbb{C}}}}

\newcommand{\Z}{\mathbb{Z}}

\DeclareMathOperator{\Span}{span}

\DeclareMathOperator{\tr}{tr}

\DeclareMathOperator{\rank}{rank}

\DeclareMathOperator{\ch}{ch}
\DeclareMathOperator{\td}{td}

\newcommand{\Spin}{{\mathop{\text{\textit{Spin}}}\nolimits}}

\newcommand{\Rep}[1]{\ensuremath{\mathbf{\underline{#1}}}}
\newcommand{\barRep}[1]{\ensuremath{\overline{\Rep{#1}}}}

\newcommand{\Osheaf}{\ensuremath{\mathscr{O}}}

\newcommand{\Kcone}{\ensuremath{\mathcal{K}}}

\newcommand{\tmod}{~\mathrm{mod}~}

\newcommand{\conditionallyincludegraphics}[2][]{%
  \IfFileExists{#2}{%
    \includegraphics[#1]{#2}%
  }{%
    \vspace{1cm}Figure \url{#2} should be here}}

\def\Vvis{V^{(1)}}
\def\Vhid{V^{(2)}}
\renewcommand{\(}{\left(}
\renewcommand{\)}{\right)}

\begin{document}

\begin{titlepage}
  \vspace*{-2cm}
  \VersionInformation
  \hfill
  \parbox[c]{5cm}{
    \begin{flushright}
    \end{flushright}
  }
  \vspace*{2cm}
  \begin{center}
    \Huge 
    Supersymmetric Hidden Sectors for Heterotic Standard Models
  \end{center}
  \vspace*{8mm}
  \begin{center}
    \begin{minipage}{\textwidth}
      \begin{center}
        \sc 
        Volker Braun${}^1$, Yang-Hui He${}^2$, and Burt A.\ Ovrut${}^3$
      \end{center}
      \begin{center}
        \textit{
          \llap{${}^1$}Dublin Institute for Advanced Studies\\
          10 Burlington Road\\
          Dublin 4, Ireland
        }
      \end{center}
      \begin{center}
        \textit{
          \llap{${}^2$}Department of Mathematics, City University\\
          London, EC1V 0HB, U.K.\\[1ex]
          School of Physics, NanKai University\\
          Tianjin, 300071, P.R. China\\
          Merton College, University of Oxford, OX1 4JD, U.K.
        }
      \end{center}
      \begin{center}
        \textit{
          \llap{${}^3$}Department of Physics\\
          University of Pennsylvania\\
          Philadelphia, PA 19104--6396
        }
      \end{center}
    \end{minipage}
  \end{center}
  \vspace*{\stretch1}
  \begin{abstract}
    Within the context of the weakly coupled $E_{8} \times E_{8}$
    heterotic string, we study the hidden sector of heterotic standard
    model compactifications to four-dimensions. Specifically, we
    present a class of hidden sector vector bundles---composed of the
    direct sum of line bundles only---that, together with an effective
    bulk five-brane, renders the heterotic standard model entirely
    $N=1$ supersymmetric. Two explicit hidden sectors are constructed
    and analyzed in this context; one with the gauge group $E_{7}
    \times U(1)$ arising from a single line bundle and a second with
    an $SO(12) \times U(1) \times U(1)$ gauge group constructed from
    the direct sum of two line bundles. Each hidden sector bundle is
    shown to satisfy all requisite physical constraints within a
    finite region of the K\"ahler cone.  We also clarify that the
    first Chern class of the line bundles need not be even in our
    context, as has often been imposed in the model building
    literature.
  \end{abstract}
  \vspace*{\stretch1}
  \begin{minipage}{\textwidth}
    \underline{\hspace{5cm}}\\
    \footnotesize
    Email: \texttt{vbraun@stp.dias.ie},
    \texttt{yang-hui.he@merton.ox.ac.uk},
    \texttt{ovrut@elcapitan.hep.upenn.edu}
  \end{minipage}
\end{titlepage}

\tableofcontents


\section{Introduction}

The ten-dimensional theory of the massless modes of weakly coupled
$E_8 \times E_8$ heterotic theory can arise in two ways. The first is
directly from the $E_8 \times E_8$ heterotic superstring after
decoupling the heavy string modes~\cite{Gross:1985fr,
  Gross:1985rr}. The second follows from compactifying
eleven-dimensional $M$-theory on an $S^{1}/\Z_{2}$ orbifold in the
limit of small radius~\cite{Horava:1995qa, Horava:1996ma,
  Lukas:1997fg, Lukas:1998ew, Lukas:1998hk, Lukas:1998tt,
  Lukas:1998yy}. Either way, the ten-dimensional effective action is a
$N=1$ supersymmetric theory with a metric, dilaton, two-form and $E_8
\times E_8$ gauge fields as the bosonic components. In addition, it
can contain topological five-branes. This can be reduced to an
$N=1$ supersymmetric theory in four dimensions by appropriately
compactifying on a smooth Calabi-Yau threefold supporting gauge fields
satisfying the hermitian Yang-Mills equations~\cite{MR86h:58038,
  MR88i:58154} with the five-branes wrapped on holomorphic
curves~\cite{Witten:1996mz}. The choice of the Calabi-Yau manifold and
the gauge field background, that is, a slope-stable holomorphic vector
bundle with vanishing slope, determines the low energy gauge group,
spectrum, and coupling parameters~\cite{Greene:1986ar, Greene:1986bm,
  Greene:1986jb, Matsuoka:1986vg, Greene:1987xh}.

In~\cite{Braun:2005nv}, it was shown that the low energy spectrum of a
specific Calabi-Yau threefold and holomorphic vector bundle with
structure group $SU(4) \subset E_8$ is exactly that of the minimal
supersymmetric standard model (MSSM) with three right-handed neutrinos
and one pair of Higgs-Higgs conjugate chiral multiplets. There are no
exotic or vector-like pairs of superfields. It was demonstrated in
\cite{Braun:2006ae} that over a specific subspace of the K\"ahler
cone, this vector bundle is slope-stable with vanishing slope and,
hence, the low energy theory is $N=1$ supersymmetric. For these
reasons, this vacuum of the \emph{observable} $E_8$ sector of the
theory was called the heterotic standard model. To complete the
vacuum, it is essential to present the explicit vector bundle for the
second $E_8$ \emph{hidden} sector. In previous work
\cite{Braun:2006me, Gray:2007zza, Gray:2007qy}, it was expedient to
choose this bundle to be trivial, requiring a five-brane sector with
non-effective cohomology class to satisfy the anomaly constraint. This
\emph{anti-brane} allows one to raise the potential energy of the
vacuum from negative to a small, positive cosmological constant
\cite{Braun:2006th}, that is, the heterotic equivalent of the KKLT
mechanism \cite{Kachru:2003aw}. It was shown in~\cite{Braun:2006ae}
that, for vanishing five-brane class, a hidden sector $SU(n)$ bundle
satisfying the Bogomolov bound~\cite{MR522939} could exist in a
subspace intersecting the region of stability of the observable sector
vector bundle. The Bogomolov bound is a necessary condition for the hidden bundle to
be $N=1$ supersymmetric. At least on certain manifolds, satisfying this bound is also
sufficient~\cite{Douglas:2006jp, Andreas:2010hv,
  Andreas:2011zs}. However, an explicit example was never constructed.

In a series of papers, both in weakly
coupled~\cite{Blumenhagen:2005ga, Blumenhagen:2005zg,
  Blumenhagen:2006ux, Blumenhagen:2006wj, Weigand:2006yj} and strongly
coupled~\cite{Lukas:1998tt, Lukas:1998hk, Anderson:2009nt} $E_8 \times
E_8$ heterotic theory, the first non-trivial corrections---string
one-loop and $M$-theory order $\kappa^{4/3}$ respectively---beyond tree level
were constructed. These were presented for both the Fayet-Iliopoulos
(FI) terms associated with anomalous $U(1)$ gauge factors and for the
gauge threshold corrections. Furthermore, it was emphasized
in~\cite{Blumenhagen:2005ga, Blumenhagen:2005zg, Blumenhagen:2006ux,
  Blumenhagen:2006wj, Weigand:2006yj} that gauge bundles with $U(n)$
structure groups, in addition to $SU(n)$ bundles, should be more
closely scrutinized. Similarly, the appearance of line bundles on the
stability wall boundaries of $SU(n)$ bundles was recognized
in~\cite{Anderson:2009nt, Anderson:2010tc}, and applied more widely to
construct new realistic models in~\cite{Anderson:2011ns,
  Anderson:2012yf, Anderson:2011vy}. In this paper, within the context
of the weakly coupled $E_8 \times E_8$ heterotic string, we will apply
these ideas to construct a completely $N=1$ supersymmetric hidden
sector for the heterotic standard model. This hidden sector will have
a non-trivial, but effective, five-brane class and a hidden sector
vector bundle composed of a direct sum of line bundles. Together with
the observable sector $SU(4)$ vector bundle, this provides an explicit
$N=1$ compactification vacuum for the heterotic standard model.

Specifically, we will do the following. In \autoref{sec:vaccum}, we
review the Calabi-Yau threefold and observable sector gauge bundle of
the heterotic standard model introduced in~\cite{Braun:2005nv,
  Braun:2006ae}. The generic form of the hidden sector bundle as a sum
of line bundles is presented, as is an analysis of the explicit
embedding of a line bundle into an $E_8$ vector bundle. Five-branes
are also briefly discussed. \autoref{sec:constraints} consists of an
analysis of the four constraint conditions required for our
compactification vacuum; namely, anomaly cancellation, slope-stability
of the observable sector bundle, $N=1$ supersymmetry of the hidden
sector bundle and positivity of the threshold corrected gauge
parameters. The explicit constraint equations for each of these
conditions are given. Two specific examples are then presented in
\autoref{sec:example}. The first is obtained using a single line bundle as
the hidden sector bundle. Its exact embedding into the hidden $E_8$ is
discussed, and it is shown to satisfy all required constraint conditions in a region of the K\"ahler cone. The associated low energy gauge group and spectrum of the hidden
sector is presented. As a second example, we choose the hidden sector bundle to be a sum of two line bundles. Again, we elucidate its exact embedding into the hidden $E_{8}$, show that it satisfies all necessary constraints in a region of the K\"ahler cone and discuss the associated low energy gauge group. Finally, important technical issues regarding the embedding of
line bundles into an $E_8$ vector bundle are discussed in
\autoref{sec:linebundles}.

The analysis in this paper has a direct, but non-trivial, extension to
strongly coupled heterotic $M$-theory~\cite{Horava:1995qa,
  Lukas:1998tt, Horava:1996ma, Lukas:1998yy, Donagi:1998xe,
  Donagi:1999gc, Donagi:1999ez, Donagi:2003tb, Braun:2004xv,
  Donagi:2004ub}. This will be presented in a future
publication.

\section{The Compactification Vacuum}
\label{sec:vaccum}

A four-dimensional, $N=1$ supersymmetric effective theory of the
weakly coupled $E_8 \times E_8$ heterotic string is obtained as
follows. First, one compactifies ten-dimensional spacetime on an
appropriate Calabi-Yau threefold, $X$. Second, it is necessary to construct
two $E_8$ gauge bundles over the Calabi-Yau
manifold. Two popular methods are to utilize a slope-stable holomorphic
vector bundle with vanishing first Chern class or a line bundle. The
latter is automatically slope-stable, but imposes the additional
physical constraint that the FI-term must vanish. The associated gauge
symmetries, spectrum and coupling parameters of the four-dimensional
theory depend on the specific choice of the compactification. In
this paper, we will examine a physically relevant subset of such
vacua; namely, those with the Calabi-Yau threefold and observable
sector vector bundle of the heterotic standard
model~\cite{Braun:2005nv, Braun:2006ae}.

\subsection{The Calabi-Yau Threefold} 

The Calabi-Yau manifold $X$ is chosen to be a torus-fibered threefold
with fundamental group $\pi_{1}(X)=\Z_3 \times \Z_3$. Specifically, it
is a fiber product of two rational elliptic $d\mathbb{P}_{9}$
surfaces, that is, a self-mirror Schoen threefold~\cite{MR923487,
  Ovrut:2002jk, Braun:2004xv, Braun:2007tp, Braun:2007xh,
  Braun:2007vy} quotiented with respect to a freely acting $\Z_3
\times \Z_3$ isometry. Its Hodge data is $h^{1,1}=h^{1,2}=3$ and,
hence, there are three K\"ahler and three complex structure
moduli. The complex structure moduli will play no role in the present
paper and we will ignore them. Relevant here is the degree-two
Dolbeault cohomology group
\begin{equation}
  H^{1,1}\big(X,\C\big)=
  \Span_{\C} \{ \omega_{1},~\omega_{2},~\omega_3 \} ,
  \label{1}
\end{equation}
where $\omega_{i}=\omega_{ia {\bar{b}}} dz^a d\bar{z}^{\bar{b}}$ are
dimensionless harmonic $(1,1)$-forms on $X$ with the property
\begin{equation}
  \omega_3\wedge\omega_3=0
  ,\quad
  \omega_{1}\wedge\omega_3=
  3\omega_{1}\wedge\omega_{1}
  ,\quad
  \omega_{2}\wedge\omega_3=
  3\omega_{2}\wedge\omega_{2}  
  .
  \label{2}
\end{equation}
Defining the intersection numbers as
\begin{equation}
  d_{ijk}=\frac{1}{v}\int_X{\omega_{i} \wedge \omega_{j} \wedge
    \omega_{k}}
  , \quad i,j,k=1,2,3
  \label{3}
\end{equation}
where $v$ is a reference volume of dimension $(\mathit{length})^6$,
it follows from \eqref{2} that
\begin{equation}
  \label{4}
  d_{ijk} = 
  \begin{pmatrix}
    \(0,\tfrac13,0\) & \(\tfrac13,\tfrac13,1\) & \(0,1,0\) \\[1mm]
    \(\tfrac13,\tfrac13,1\) & \(\tfrac13,0,0\) & \(1,0,0\) \\[1mm]
    \(0,1,0\) & \(1,0,0\) & \(0,0,0\)
  \end{pmatrix} 
  .
\end{equation}
The $\{ij\}$-th entry in the matrix corresponds to the triplet
$(d_{\{ij\}k}| k=1,2,3)$.

Our analysis will require the Chern classes of the tangent bundle
$TX$. Noting that the associated structure group is $SU(3) \subset
SO(6)$, it follows that ${\rm rank}(TX)=3$ and
$c_{1}(TX)=0$. Furthermore, the self-mirror property of this specific
threefold implies $c_3(TX)=0$. Finally, we find that
\begin{equation}
  c_{2}(TX)=\frac{1}{v^{2/3}}
  \big(
  12 \omega_{1}\wedge\omega_{1}+12\omega_{2}\wedge\omega_{2}
  \big) .
  \label{5}
\end{equation}
We will use the fact that if one chooses the generators of $SU(3)$ to
be hermitian, then the second Chern class of the tangent bundle can be
written as
\begin{equation}
  c_{2}(TX)  =
  -\frac{1}{16\pi^{2}}\tr_{SO(6)} R \wedge R ,
  \label{6}
\end{equation}
where $R$ is the Lie algebra valued curvature two-form.

Note that $H^{2,0}=H^{0,2}=0$ on a Calabi-Yau threefold. It follows
that $H^{1,1}(X,\C)=H^{2}(X,\R)$ and, hence, $\omega_{i}$, $i=1,2,3$
span the real vector space $H^{2}(X,{\R})$. Furthermore, it was shown
in~\cite{Gomez:2005ii} that the curve Poincar\'e dual to each two-form
$\omega_{i}$ is effective. Therefore, the K\"ahler cone is the
positive octant
\begin{equation}
  {\cal{K}}=H^{2}_{+}(X,{\R})
  \subset H^{2}(X,{\R}) .
  \label{7}
\end{equation}
The K\"ahler form, defined to be $\omega_{a {\bar{b}}}=ig_{a
  {\bar{b}}}$ where $g_{a {\bar{b}}}$ is the Calabi-Yau metric, can be
any element of ${\cal{K}}$. That is, the K\"ahler form can be expanded
as
\begin{equation}
  \omega=a^{i}\omega_{i} , \quad a^{i} >0, \quad i=1,2,3 .
  \label{8}
\end{equation}
The real, positive coefficients $a^{i}$ are the three $(1,1)$ K\"ahler
moduli of the Calabi-Yau threefold. Here, and through this paper,
upper and lower $H^{1,1}$ indices are summed unless otherwise
stated. The dimensionless volume modulus is defined by
\begin{equation}
V=\frac{1}{v}\int_{X}{\sqrt{^{6}g}}
\label{9}
\end{equation}
and, hence, the dimensionful Calabi-Yau volume is $vV$. Using the
definition of the K\"ahler form and \eqref{3}, $V$ can be written as
\begin{equation}
  V=\frac{1}{6v}\int_{X}{\omega \wedge \omega \wedge \omega}=
  \frac{1}{6}d_{ijk}a^{i}a^{j}a^{k} .
  \label{10}
\end{equation}
It is useful to express the three $(1,1)$ moduli in terms of $V$ and
two additional independent moduli. This can be accomplished by
defining the scaled shape moduli
\begin{equation}
  b^{i}=V^{-1/3}a^{i} , \quad i=1,2,3 .
  \label{11}
\end{equation}
It follows from \eqref{10} that they satisfy the constraint
\begin{equation}
  d_{ijk}b^{i}b^{j}b^{k}=6
  \label{12}
\end{equation}
and, hence, represent only two degrees of freedom. Finally, note that
all moduli defined thus far, that is, $a^{i}$, $V$ and $b^{i}$, are
functions of the four coordinates $x^{\mu}$, $\mu=0,\dots,3$ of
Minkowski space $M_{4}$.

\subsection{The Observable Sector Gauge Bundle} 

The $E_8 \times E_8$ vector bundle $V$ over $X$ is a direct sum of
an \emph{observable} sector bundle, $V^{(1)}$, whose structure group
is embedded in the first $E_8$ factor, with a \emph{hidden} sector
bundle, $V^{(2)}$, with structure group in the second
$E_8$. $V^{(1)}$ is chosen to be holomorphic with structure group
$SU(4)\subset E_8$, thus breaking
\begin{equation}
  E_8 \longrightarrow Spin(10) .
  \label{13}
\end{equation}
Our analysis will require the Chern classes of $V^{(1)}$. Since the
structure group is $SU(4)$, it follows immediately that ${\rm
  rank}(V^{(1)})=4$ and $c_{1}(V^{(1)})=0$. The heterotic standard
model is constructed to have the observed three chiral families of
quarks/leptons and, hence, $V^{(1)}$ must be chosen so that
$c_3(V^{(1)})=3$. Finally, we find that
\begin{equation}
  c_{2}(V^{(1)})=
  \frac{1}{v^{2/3}}\big(\omega_{1} \wedge \omega_{1}+
  4~\omega_{2} \wedge \omega_{2}+ 4~\omega_{1} \wedge \omega_{2}\big) .
\label{14}
\end{equation}
Here, and below, it will be useful to note the following. Let
${\cal{V}}$ be an arbitrary vector bundle with structure group ${\cal{G}}$,
and ${\cal{F}^{\cal{V}}}$ the associated Lie algebra valued two-form
gauge field strength. If the generators of ${\cal{G}}$ are chosen to be
hermitian, then
\begin{equation}
  \frac{1}{8\pi^{2}}
  \tr_{\cal{G}} {\cal{F}^{\cal{V}}} \wedge {\cal{F}}^{\cal{V}}
  =\ch_{2}({\cal{V}})
  =\frac{1}{2}c_{1}({\cal{V}}) 
  \wedge c_{1}({\cal{V}})-c_{2}({\cal{V}}) ,
\label{15}
\end{equation}
where $\ch_{2}({\cal{V}})$ is the second Chern character of
${\cal{V}}$. Furthermore, we denote by $\tr_{\cal{G}}$ the trace in the fundamental
representation of the structure group ${\cal{G}}$ of the bundle. When applied
to the vector bundle $V^{(1)}$ in the observable sector, it follows
from $c_{1}(V^{(1)})=0$ that
\begin{equation}
  c_{2}(V^{(1)})
  =
  -\frac{1}{8\pi^{2}}\tr_{SU(4)} F^{(1)} \wedge F^{(1)} 
  =
  -\frac{1}{16\pi^{2}}\tr_{E_8} F^{(1)} \wedge F^{(1)} 
  ,
  \label{16}
\end{equation}
where $F^{(1)}$ is the gauge field strength for the visible sector
bundle $V^{(1)}$ and $\tr_{E_{8}}$ indicates the trace is over the fundamental $\Rep{248}$ representation of $E_{8}$. Note that the conventional normalization of the
trace $\tr_{E_8}$ includes a factor of
$\tfrac{1}{30}$, the inverse of the dual Coxeter number of $E_8$. We have expressed $c_{2}(V^{(1)})$ in terms of $\tr_{E_{8}}$ since the fundamental $SU(4)$ representation must be embedded into the adjoint representation of $E_{8}$ in the observable sector. 

To preserve $N=1$ supersymmmetry in four-dimensions, $V^{(1)}$ must be
both slope-stable and have vanishing slope~\cite{Green:1987sp,
  Green:1987mn}. In the context of this paper, these constraints are
most easily examined in the $d=4$ effective theory and, hence, will be
discussed below.  Finally, when two \emph{flat} Wilson lines are
turned on, each generating a different $\Z_3$ factor of the $\Z_3
\times \Z_3$ holonomy of $X$, the observable gauge group is further
broken to
\begin{equation}
  Spin(10) 
  \longrightarrow 
  SU(3)_{C} \times SU(2)_{L} \times U(1)_{Y} \times U(1)_{B-L} .
  \label{17}
\end{equation}

\subsection{ The Hidden Sector Gauge Bundle} 

In the hidden sector, the vector bundle $V^{(2)}$ introduced in this
paper will be constructed entirely as the sum of holomorphic line
bundles. Let us briefly review the properties of such bundles on our
specific geometry. Line bundles are classified by the divisors of $X$
and, hence, equivalently by the elements of the integral cohomology
\begin{equation}
  H^{2}
  (X,\Z)= 
  \big\{
  a \omega_{1}+b \omega_{2}+c \omega_3
  ~\big|~
  a,b,c \in \Z 
  ,~
  a+b = 0\tmod 3
  \big\}  .
  \label{18}
\end{equation}
It is conventional to denote the line bundle associated with the
element $a \omega_{1}+b \omega_{2}+c \omega_3$ of $H^{2}(X,\Z)$ as
\begin{equation}
  \Osheaf_X(a,b,c) .
  \label{19}
\end{equation}
Note that the $\omega_1$, $\omega_2$, $\omega_3$ are the natural basis
of invariant integral forms on the \emph{covering space}. In order to
correspond to integral forms on the quotient Calabi-Yau manifold $X$,
an element $a \omega_{1}+b \omega_{2}+c \omega_3$ has to satisfy the
additional constraint $a+b=0 \tmod 3$ in order to be integral. This
can also be seen from the intersection numbers \eqref{4}, which
are naively fractional. Only the intersection of classes satisfying
$a+b=0 \tmod 3$ is integral. For the purposes of constructing a
heterotic gauge bundle from a line bundle $\Osheaf_X(a,b,c)$, this is
the only constraint required on the integers $a,b,c$. In particular,
as explained in \autoref{sec:linebundles}, it is not necessary to
impose that they be even for there to exist a spin structure on
$\Vhid$. Although the auxiliary line bundle is not spin if $a$, $b$,
$c$ are not even, the $E_8$ bundle is always spin.

We will choose the the hidden sector bundle to be
\begin{equation}
  V^{(2)}=\bigoplus_{r=1}^{R}L_{r},~\quad L_{r}
  =\Osheaf_{X}(\ell^{1}_{r},\ell^{2}_{r},\ell^{3}_{r})
  \label{eq:V2ell}
\end{equation}
where
\begin{equation}
  \ell^{1}_{r}+\ell^{2}_{r}=0~{\rm mod}~3
  ,\quad
  r=1,\dots,R 
  \label{22}
\end{equation}
for some positive integer $R$. The structure group is $U(1)^{R}$, where
each $U(1)$ factor has a specific embedding into the hidden sector
$E_8$ gauge group. It follows from the definition that
$\rank(V^{(2)})=R$ and that the first Chern class is
\begin{equation}
  c_{1}(V^{(2)})=\sum_{r=1}^{R}c_{1}(L_{r})
  ,\quad
  c_{1}(L_{r})=\frac{1}{v^{1/3}}
  \big(
  \ell^{1}_{r} \omega_{1}+\ell^{2}_{r}\omega_{2}+\ell^{3}_{r}\omega_3
  \big) .
\label{23}
\end{equation}
Note that since $V^{(2)}$ is a sum of holomorphic line bundles,
$c_{2}(V^{(2)})=c_3(V^{(2)})=0$. However, the relevant quantity for
the hidden sector vacuum is related to the second Chern character given in
\eqref{15}. Defining $F^{(2)}$ to be the gauge field strength for the
hidden sector bundle $V^{(2)}$, this becomes
\begin{equation}
  \frac{1}{8\pi^{2}}\tr_{U(1)^R} F^{(2)} \wedge F^{(2)} 
  =
  \ch_{2}(V^{(2)})
  =
  \sum_{r=1}^{R} \ch_{2}(L_{r})
  =
  \sum_{r=1}^{R} \frac{1}{2} c_1(L_r)\wedge c_1(L_r)
  \label{24}
\end{equation}
since $c_{2}(L_{r})=0$. As in the observable sector, the $U(1)^{R}$  fundamental representation must be embedded into the adjoint representation of the hidden sector $E_{8}$. Hence, the physically relevant quantity is proportional to $\tr_{E_8}$ of $F^{(2)} \wedge F^{(2)}$, where we remind the reader that
our normalization of the $E_8$ trace includes the $\tfrac{1}{30}$
as in \eqref{16}. Specifically, the term of interest is
\begin{equation}
  \frac{1}{16\pi^{2}}\tr_{E_8} F^{(2)} \wedge F^{(2)} 
  =
  \sum_{r=1}^{R}
  \frac{2a_r}{8\pi^{2}}\tr_{U(1)} F^{(2)}_r \wedge F^{(2)}_r
  =
  \sum_{r=1}^{R} a_r\, c_1(L_r)^2
  \label{25}
\end{equation}
with a group-theoretic factor
\begin{equation}
  a_{r}=\frac{1}{4}\tr_{E_8}Q_{r}^{2}
  ,
  \label{26}
\end{equation}
where $Q_{r}$ is the generator of the $r$-th $U(1)$ factor embedded
into the $\Rep{248}$ representation of the hidden sector $E_8$. Note that we have used \eqref{24} in going from the second to the third term in \eqref{25}.

The definition of $a_r$ with the coefficient $\tfrac{1}{4}$ is, of
course, a convention. However, it is justified by the following
computation which we leave as an exercise for the reader to verify.
Using the embedding $U(1)\subset SU(2) \subset E_7$ defined by
eqns.~\eqref{eq:U1SU2} and~\eqref{73} below, the normalized trace is given by
$\tr_{E_8}Q^2 = \tfrac{1}{30} \cdot 60 \cdot 2$. Therefore, the
minimal $U(1)$ embedding in $E_8$ leads to $a_{\min}=1$, explaining
the conventional normalization factor of $\tfrac{1}{4}$ in
\eqref{26}. In fact, by comparing the usual formula for the Chern
character of a line bundle \eqref{24} with the $E_8$
characteristic class \eqref{25}, one might have guessed that $a_r$
is half-integral. However, this is not true and $a_r$ is always
integral. This is also crucially important for the contribution to the
heterotic anomaly, which must be an integral cohomology class, to be
well-defined. In \autoref{sec:linebundles}, we will discuss this in
more detail.

\subsection{Wrapped Five-Branes}

In addition to the holomorphic vector bundles in the observable and
hidden sectors, the compactification can also contain five-branes
wrapped on two-cycles ${\cal{C}}_{2}^{(n)}$, $n=1,\dots,M$ in
$X$. Cohomologically, each such five-brane is described by the
(2,2)-form Poincar\'e dual to ${\cal{C}}_{2}^{(n)}$, which we denote by
$W^{(n)}$. Note that to preserve $N=1$ supersymmetry in the
four-dimensional theory, these curves must be holomorphic and, hence,
each $W^{(n)}$ be an effective class.

\section{The Vacuum Constraint Conditions}
\label{sec:constraints}

In order for the Calabi-Yau threefold $X$, the observable and hidden
sector vector bundles $V^{(1)}$, $V^{(2)}$ and the five-branes
$W^{(n)}$ discussed above to form a consistent compactification, they
must satisfy a set of physical constraints. These are the following.

\subsection{Anomaly Cancellation} 

As discussed in~\cite{Green:1987sp, Green:1987mn, Lukas:1998tt,
  Lukas:1998hk, Lukas:1998yy}, anomaly cancellation requires that
\begin{equation}
  -\frac{1}{16 \pi^{2}} \tr_{SO(6)} R \wedge R
  +\frac{1}{16 \pi^{2}} \tr_{E_{8}} F^{(1)} \wedge F^{(1)}
  +\frac{1}{16 \pi^{2}} \tr_{E_{8}} F^{(2)} \wedge F^{(2)}
  -\sum_{m=1}^{M}W^{(m)}=0 .
  \label{28}
\end{equation}
Using \eqref{6},\eqref{16} and \eqref{24},\eqref{25} the anomaly
cancellation condition can be expressed as
\begin{equation}
  c_{2}(TX)-c_{2}(V^{(1)})+
  \sum_{r=1}^{R}a_{r}c_{1}(L_{r})\wedge c_{1}(L_{r})-W=0 ,
  \label{29}
\end{equation}
where $W=\sum_{m=1}^{M}W^{(m)}$.

Condition \eqref{29} is expressed in terms of four-forms in
$H^{4}(X,\R)$. We find it easier to analyze its consequences by
writing it in the dual homology space $H_{2}(X,\R)$. In this case, the
coefficient of the $i$-th vector in the basis dual to
$(\omega_{1},\omega_{2},\omega_3)$ is given by wedging each term in
\eqref{29} with $\omega_{i}$ and integrating over $X$. Using
\eqref{5},\eqref{14} and the intersection numbers \eqref{3},\eqref{4}
gives
\begin{equation}
  \frac{1}{v^{1/3}}\int_{X}{\Big(c_{2}(TX)-c_{2}(V^{(1)}) \Big)
    \wedge \omega_{1,2,3}}
  =\big(\tfrac{4}{3},\tfrac{7}{3},-4\big) .
\label{30}
\end{equation}
Similarly, \eqref{3},\eqref{4} and \eqref{23} imply
\begin{equation}
  \frac{1}{v^{1/3}}\int_{X}{c_{1}(L_{r}) 
    \wedge c_{1}(L_{r})\wedge \omega_{i}}
  =  d_{ijk} \ell^j_r \ell^k_r, \quad i=1,2,3 .
  \label{31}
\end{equation}
Defining
\begin{equation}
  W_{i}=\frac{1}{v^{1/3}}\int_{X}{W \wedge \omega_{i}} ,
  \label{32}
\end{equation}
it follows that the anomaly condition \eqref{29} can be expressed as 
\begin{equation}
  W_{i} = 
  \big(\tfrac{4}{3},\tfrac{7}{3},-4\big)_{i}
  +\sum_{r=1}^{R} a_{r} d_{ijk} \ell^j_r \ell^k_r \geq 0
  , \quad i=1,2,3  .
  \label{33}
\end{equation}
The semi-positivity constraint on $W$ follows from the requirement
that it be an effective class to preserve $N=1$ supersymmetry.

\subsection{Slope-Stability of the Observable Sector Bundle}

As mentioned previously, to preserve $N=1$ supersymmmetry in
four-dimensions the holomorphic $SU(4)$ vector bundle $V^{(1)}$
associated with the observable $E_8$ gauge group must be both
\begin{itemize}
\item slope-stable (to admit a solution to the Hermitian Yang-Mills
  equation), and
\item have vanishing slope (because there is no FI term for $SU(n)$
  bundles).
\end{itemize}
Here, the slope of any bundle or sub-bundle $\cal{F}$ is defined as
\begin{equation}
  \mu({\cal{F}})=\frac{1}{\rank({\cal{F}})v^{2/3}} 
  \int_{X}{c_{1}(\cal{F})\wedge \omega \wedge \omega} ,
  \label{50}
\end{equation}
where $\omega=a^i\omega_i$ is the K\"ahler form as in \eqref{8}. The
rank-4 bundle $V^{(1)}$ has vanishing slope since
$c_{1}(V^{(1)})=0$. But, is it slope-stable? As shown in detail
in~\cite{Braun:2006ae, Braun:2006th}, one can identify a set of $7$
``maximally destabilizing'' line sub-bundles 
\begin{equation}
  \begin{gathered}
    \Osheaf_X(1, -1, -1),~
    \Osheaf_X(-1, 1, -1),~
    \Osheaf_X(-2, 2, 0),~
    \Osheaf_X(2, -2, -1),~
    \\
    \Osheaf_X(2, -5, 1),~
    \Osheaf_X(1, -4, 1),~
    \Osheaf_X(-4, 1, 1).
  \end{gathered}
\end{equation}
It is a sufficient condition for stability to have all of their slopes
be negative simultaneously. This singles out the subspace of the
K\"ahler cone defined by the following 7 inequalities
\begin{equation}
  \label{51}
  \begin{split}
    -3 (a^1 - a^2) (a^1 + a^2 + 6 a^3) - 18 a^1 a^2 
    <&\; 0\\
    3 (a^1 - a^2) (a^1 + a^2 + 6 a^3) - 18 a^1 a^2 
    <&\; 0\\
    6 (a^1 - a^2) (a^1 + a^2 + 6 a^3)
    <&\; 0\\
    -6 (a^1 - a^2) (a^1 + a^2 + 6 a^3) - 18 a^1 a^2 
    <&\; 0\\
    -3 (5 a^1 - 2 a^2) (a^1 + a^2 + 6 a^3) + 9 a^1 a^2 
    <&\; 0\\
    -3 (4 a^1 - a^2) (a^1 + a^2 + 6 a^3) + 9 a^1 a^2 
    <&\; 0\\
    3 (a^1 - 4 a^2) (a^1 + a^2 + 6 a^3) + 9 a^1 a^2 
    <&\; 0 .
  \end{split}
\end{equation}
These can be slightly simplified into the statement that $a^{i}$,$i=1,2,3$ must satisfy at least one of the two sets of inequalities
\newcommand{\VisibleStabilityCond}{
  \begin{gathered}
   \left(
    a^1
    < 
    a^2
    \leq 
    \sqrt{\tfrac{5}{2}} a^1
    \quad\text{and}\quad
    a^3
    <
    \frac{
      -(a^1)^2-3 a^1 a^2+ (a^2)^2
    }{
      6 a^1-6 a^2
    } 
    \right)
  \quad\text{or}\\
  \left(
    \sqrt{\tfrac{5}{2}} a^1
    <
    a^2
    <
    2 a^1
    \quad\text{and}\quad
    \frac{
      2(a^2)^2-5 (a^1)^2
    }{
      30 a^1-12 a^2
    }
    <
    a^3
    <
    \frac{
      -(a^1)^2-3 a^1 a^2+ (a^2)^2
    }{
      6 a^1-6 a^2
    }
   \right) 
  \end{gathered}
 }
\begin{equation}
  \VisibleStabilityCond
  \label{eq:KsCond}
\end{equation}
The subspace $\Kcone^{s}$ satisfying \eqref{51} is a full-dimensional
\begin{figure}[htb]
  \centering
  \conditionallyincludegraphics[width=0.9\textwidth]{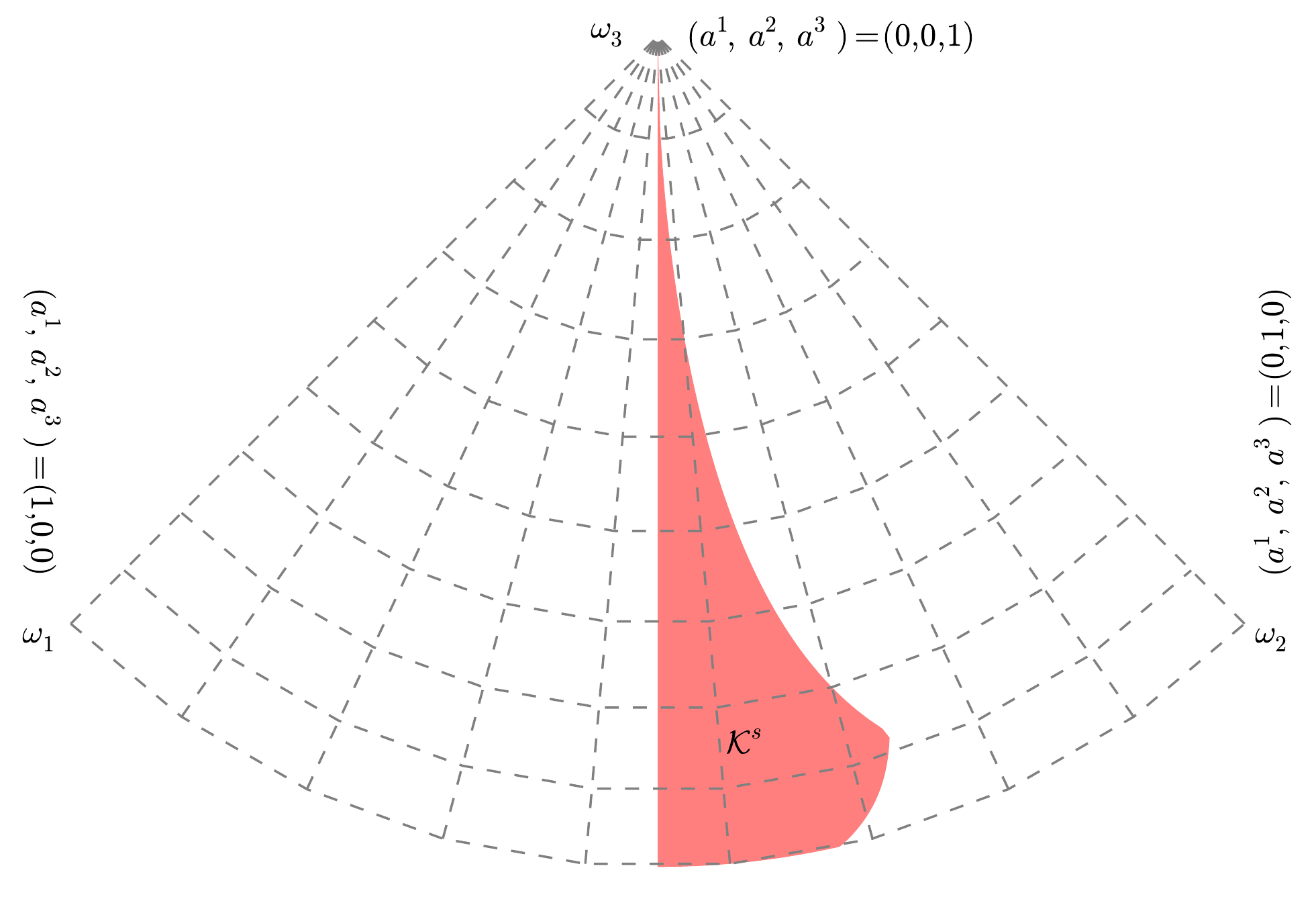}
  \caption{Map projection of the unit sphere intersecting the K\"ahler
    cone, that is, the positive octant in $H^2\big(X,\R\big)\simeq
    \R^3$. The visible sector bundle $\Vvis$ is stable inside the red
    teardrop-shaped region $\Kcone^s$. Every point in the projection
    represents a ray in the K\"ahler cone. For example,
    $(a^{1},a^{2},a^{3})=(0,1,0)$ generates the ray in the
    $\omega_{2}$ direction.  }
  \label{fig:starmap}
\end{figure}
subcone of the K\"ahler cone $\Kcone$ defined in \eqref{7}. It is a
cone because the inequalities are homogeneous. In other words, only
the angular part of the K\"ahler moduli are constrained but not the
overall volume. Hence, it is best displayed as a two-dimensional
``star map'' as seen by an observer at the origin. This is shown in
\autoref{fig:starmap}. For K\"ahler moduli restricted to this subcone,
the four-dimensional low energy theory in the observable sector is
$N=1$ supersymmetric.

\subsection
[{\texorpdfstring{$N=1$}{N=1} 
  Supersymmetric Hidden Sector Bundle}]
{\texorpdfstring{\boldmath$N=1$}{N=1} 
  Supersymmetric Hidden Sector Bundle}

In the heterotic standard model vacuum, the observable sector vector
bundle $V^{(1)}$ has structure group $SU(4)$. Hence, it does not lead
to an anomalous $U(1)$ gauge factor in the observable sector of the
low energy theory. However, the hidden sector bundle $V^{(2)}$
introduced above consists entirely of a sum of line bundles and,
therefore, has structure group $U(1)^{R}$. Each $U(1)$ factor leads to
an anomalous $U(1)$ gauge group in the four-dimensional effective
field theory and, hence, an associated $D$-term.

Let $L_{r}$ be any one of the sub-line bundles of $V^{(2)}$. Then, it
was shown in~\cite{Blumenhagen:2005ga} that the associated FI term is
\begin{multline}
  \label{56}
  FI^{U(1)_{r}} 
  ~\propto~
  \mu(L_{r}) - 
  \frac{g_{s}^{2}l_{s}^{4}}{v^{2/3}} 
  \int_{X} c_{1}(L_{r})\wedge 
  \\
  \left(\sum_{s=1}^{R} a_{s} c_{1}(L_{s}) \wedge c_{1}(L_{s}) 
  + \frac{1}{2} c_{2}(TX)-\sum_{m=1}^{M} 
  (\tfrac{1}{2}+\lambda_{m})^{2} W^{(m)} \right)
  ,    
\end{multline}
where $\mu(L_{r})$ is given in \eqref{50} and
\begin{equation}
  g_{s}=e^{\phi_{10}}
  ,\quad
  l_{s}=2\pi\sqrt{\alpha^{\prime}} 
  \label{56A}
\end{equation}
are the string coupling and string length respectively. Furthermore,
each $\lambda_{n}$ is a real modulus that, together with a self-dual
two-form ${\tilde{B}}_{n}$, forms a tensor multiplet on the
six-dimensional worldvolume of the $n$-th five-brane. The
normalization of these moduli is chosen so that
\begin{equation}
  \lambda_{n}\in \left[
    -\tfrac{1}{2}, \tfrac{1}{2}
  \right]
  .
  \label{56B}
\end{equation}
The first term on the right-hand side, that is, the slope of $L_{r}$
defined in \eqref{50}, is the tree level result. 

The remaining terms are the string one-loop string corrections first
presented in~\cite{Blumenhagen:2005ga, Blumenhagen:2005pm,
  Blumenhagen:2006wj}. The general form of each $D$-term is as the sum
of 1) the moduli dependent FI parameter \eqref{56} and 2) terms
quadratic in the fields charged under the gauge symmetry weighted by
their specific charge. In this paper, for simplicity, we will assume
that all $U(1)^{R}$ charged zero-modes in the hidden sector have
vanishing expectation values. Then the hidden sector will be $N=1$
supersymmetric if and only if the moduli-dependent FI parameter
vanishes for each $L_{r}$. That is,
\begin{multline}
  \int_{X}{c_{1}(L_{r}) \wedge \omega \wedge
    \omega}-g_{s}^{2}l_{s}^{4}  
  \int_{X} c_{1}(L_{r}) \wedge
  \\
  \left(\sum_{s=1}^{R} a_{s} c_{1}(L_{s}) \wedge c_{1}(L_{s}) 
  +\frac{1}{2} c_{2}(TX) -\sum_{m=1}^{M} (\tfrac{1}{2}+\lambda_{m})^{2}
  W^{(m)} \right)  = 0  
  \label{57}
\end{multline}
for $r=1,\dots,R$. Using \eqref{3}, \eqref{4}, \eqref{8}, \eqref{23},
\eqref{31}, \eqref{32} and noting from \eqref{5} that
\begin{equation}
  \frac{1}{v^{1/3}}\int_{X}{\frac{1}{2}c_{2}(TX) \wedge
    \omega_i}=(2,2,0)_i
  ,
  \label{58}
\end{equation}
it follows that for each $L_{r}$ condition \eqref{57} can be written
as
\begin{equation}
  \label{59} 
  d_{ijk}l_{r}^{i}a^{j}a^{k}- 
  \frac{g_{s}^{2}l_{s}^{4}}{v^{2/3}} 
  \left(
  d_{ijk}\ell_{r}^{i}
  \sum_{s=1}^{R}a_{s}\ell_{s}^{j}\ell_{s}^{k} 
  + \ell^{i}_{r}(2,2,0)_{i}
  -\sum_{m=1}^{M}(\tfrac{1}{2}+\lambda_{m})^{2}\ell_{r}^{i}W^{(m)}_{i}
  \right)
  = 0  
\end{equation}
where
\begin{equation}
  V=\frac{1}{6}d_{ijk}a^{i}a^{j}a^{k} 
  .
  \label{60}
\end{equation}

\subsection{Gauge Threshold Corrections}

The gauge couplings of the non-anomalous components of the $d=4$ gauge
group, in both the observable and hidden sectors, have been computed
to the string one-loop level in~\cite{Blumenhagen:2005ga,
  Blumenhagen:2005zg, Blumenhagen:2006ux, Blumenhagen:2006wj,
  Weigand:2006yj}. Including five-branes, these are given by
\begin{flalign}
  \hspace{10mm}
  \begin{split}
    \frac{4\pi}{g^{(1) 2}} = &\;
    \frac{1}{6v}\int_{X}\omega \wedge \omega \wedge \omega 
    -\frac{g_{s}^{2}l_{s}^{4}}{2v} \int_{X}\omega \wedge 
    \\&\hspace{2cm}
    \left(
      -c_{2}(V^{(1)})
      +\frac{1}{2}c_{2}(TX)
      -\sum_{m=1}^{M}(\tfrac{1}{2}-\lambda_{m})^{2}W^{(m)}  
    \right)
  \end{split}
  &
  \label{64}
\end{flalign}
and 
\begin{flalign}
  \hspace{10mm}
  \begin{split}
    \frac{4\pi}{g^{(2) 2}}=&\;
    \frac{1}{6v}\int_{X}\omega \wedge \omega \wedge \omega 
    -\frac{g_{s}^{2}l_{s}^{4}}{2v}\int_{X}\omega 
    \wedge 
    \\&\hspace{1cm}
    \left(\sum_{r=1}^{R} a_{r} c_{1}(L_{r}) \wedge c_{1}(L_{r}) 
      +\frac{1}{2}c_{2}(TX)-\sum_{m=1}^{M}(\tfrac{1}{2}+\lambda_{m})^{2}W^{(m)} 
    \right)
    ,
  \end{split}
  &
  \label{65}
\end{flalign}
respectively. The first term on the right-hand side, that is, the
volume $V$ defined in \eqref{10}, is the tree level result. The
remaining terms are the one-loop corrections first presented
in~\cite{Blumenhagen:2005ga}.

Clearly, consistency of the $d=4$ effective theory requires both
$g^{(1) 2}$ and $g^{(1) 2}$ to be positive. It follows that the moduli
of the four-dimensional theory are constrained to satisfy
\begin{flalign}
  \hspace{10mm}
  \begin{split}
    \frac{1}{3}\int_{X}\omega \wedge \omega \wedge \omega 
    -g_{s}^{2}l_{s}^{4}\int_{X}\omega \wedge
    \bigg(& -c_{2}(V^{(1)}) 
    \\[-1ex] &
    +\frac{1}{2}c_{2}(TX)-\sum_{m=1}^{M}(\tfrac{1}{2}-\lambda_{m})^{2}W^{(m)}
    \bigg)
    ~>~ 0
  \end{split}
  &
\label{66}
\end{flalign}
and 
\begin{flalign}
  \hspace{10mm}
  \begin{split}
    \frac{1}{3}\int_{X}\omega \wedge \omega \wedge \omega 
    -g_{s}^{2}l_{s}^{4}\int_{X}\omega \wedge 
    \bigg(
    & \sum_{r=1}^{R} a_{r} c_{1}(L_{r}) \wedge c_{1}(L_{r})
    \\ &
    +\frac{1}{2}c_{2}(TX)
    -\sum_{m=1}^{M}(\tfrac{1}{2}+\lambda_{m})^{2}W^{(m)}  
    \bigg) 
    ~>~ 0 .
  \end{split}
  &
\label{67}
\end{flalign}
As in the previous subsections, one can use \eqref{3}, \eqref{4},
\eqref{8}, \eqref{14}, \eqref{31}, \eqref{32} and \eqref{58} to
rewrite these expressions as
\begin{flalign}
  d_{ijk} a^i a^j a^k- 3 \frac{g_{s}^{2}l_{s}^{4}}{v^{2/3}}
  \left(
    -(\tfrac83 a^1 + \tfrac53 a^2 + 4 a^3)
    +
    2(a^1+a^2) -\sum_{m=1}^{M}(\tfrac{1}{2}-\lambda_{m})^{2} a^i \;W^{(m)}_i
  \right) > 0 
  &\label{68}
\end{flalign}
and
\begin{flalign}
  d_{ijk} a^i a^j a^k- 3 \frac{g_{s}^{2}l_{s}^{4}}{v^{2/3}}
  \left(
    d_{ijk}a^i  \sum_{r=1}^{R}a_{r} \ell^j_r \ell^k_r
    + 2(a^1+a^2) -\sum_{m=1}^{M}(\tfrac{1}{2}+\lambda_{m})^{2} a^i \;W^{(m)}_i
  \right) > 0 
  &\label{69}
\end{flalign}
respectively.

\section{Specific Examples}
\label{sec:example}

\subsection{Constraints for a Single Line Bundle}

We now present an explicit $N=1$ supersymmetric hidden sector for the
weakly coupled heterotic standard model that satisfies all vacuum
constraints. To do this, one must specify the number of line bundles
$L_{r}$ and their exact embeddings into the hidden vector bundle
$E_8$. Later in this section, we will consider the case of two
independent line bundles. However, for now we restrict ourselves to the
simplest example consisting only of a single line bundle
\begin{equation}
  V^{(2)}=L,~\quad L=\Osheaf_X(\ell^{1},\ell^{2},\ell^{3})
  \label{70}
\end{equation}
parametrized by integers
\begin{equation}
  \ell^{1},\ell^{2},\ell^{3} \in \Z
  , \quad 
  \ell^{1}+\ell^{2}=0 ~\text{mod}~3 .
  \label{71}
\end{equation}
Furthermore, the explicit embedding of $L$ into $E_8$ is chosen as
follows. First, recall that
\begin{equation}
  SU(2) \times E_{7} \subset E_8
  \label{72}
\end{equation}
is a maximal subgroup. With respect to $SU(2) \times E_7$, the
\Rep{248} representation of $E_8$ decomposes as
\begin{equation}
  \Rep{248} \longrightarrow
  (\Rep{1},\Rep{133}) 
  \oplus(\Rep{2},\Rep{56})\oplus (\Rep{3}, \Rep{1}) .
  \label{73}
\end{equation}
We embed the generator $Q$ of the U(1) structure group of $L$---more
specifically, the generator of the $S(U(1)\times U(1))$ Abelian group
of the induced rank two bundle $L \oplus L^{*}$ in $SU(2)$---so that under $SU(2)
\rightarrow U(1)$ the two-dimensional $SU(2)$ representation
decomposes as
\begin{equation}
  \Rep{2} \longrightarrow 1 \oplus -1 .
  \label{74}
\end{equation}
It follows that under $U(1) \times E_{7}$
\begin{equation}
  \Rep{248} \longrightarrow 
  (0,\Rep{133})
  \oplus \Big((1,\Rep{56}) \oplus (-1,\Rep{56})\Big) 
  \oplus \Big( (2,\Rep{1}) \oplus (0,\Rep{1}) \oplus (-2,\Rep{1}) \Big) .
  \label{75}
\end{equation}
The generator $Q$ of this embedding of the line bundle can be read off
from expression \eqref{75}. Inserting this into \eqref{26}, we find that
\begin{equation}
  a=1 .
  \label{76}
\end{equation}

Having presented the hidden sector vector bundle, one must specify the
number of five-branes. For simplicity, we assume that there is only
one five-brane in this example. It then follows from \eqref{33},
\eqref{59}, \eqref{68} and \eqref{69} that the constraints for this
explicit example are given by
\begin{subequations}
  \begin{align}
    W_{i}= 
    \big(\tfrac{4}{3},\tfrac{7}{3},-4\big)_{i}
    +d_{ijk} \ell^j \ell^k 
    &\;\geq 0
    , \quad i=1,2,3  
    \label{eq:U1_W}
    \\[1ex]
    d_{ijk}\ell^{i}a^{j}a^{k}
    - \frac{g_{s}^{2}l_{s}^{4}}{v^{2/3}}\Big( d_{ijk}\ell^{i} \ell^{j}\ell^{k} 
    + \ell^{i}(2,2,0)_{i}
    \qquad &\nonumber \\
    -\big(\tfrac{1}{2}+\lambda\big)^{2}\ell^{i}W_{i}\Big) 
    &\;= 0
    , 
    \label{eq:U1_FI} 
    \\[1ex]
    d_{ijk} a^i a^j a^k
    - 3  \frac{g_{s}^{2}l_{s}^{4}}{v^{2/3}} \Big(
    -\big(\tfrac83 a^1 + \tfrac53 a^2 + 4 a^3\big) +
    \qquad & \nonumber \\
    + 2(a^1+a^2) -\big(\tfrac{1}{2}-\lambda\big)^{2} a^i W_i\Big) 
    &\;> 0 ,
    \label{eq:U1_gvis}
    \\[1ex]
    d_{ijk} a^i a^j a^k- 3  \frac{g_{s}^{2}l_{s}^{4}}{v^{2/3}}
    \Big(
    d_{ijk}a^i \ell^j \ell^k + 
    \qquad &\nonumber \\
    + 2(a^1+a^2) -\big(\tfrac{1}{2}+\lambda\big)^{2} a^i W_i\Big) 
    &\;> 0  .
    \label{eq:U1_ghid}
  \end{align}
  These constraints have to be solved simultaneously with the
  condition \eqref{eq:KsCond} for the slope-stability of the
  observable $E_8$ non-Abelian vector bundle. That is,
  \begin{multline}
    \VisibleStabilityCond
    \label{eq:U1_visible}
  \end{multline}
\end{subequations}
We now seek simultaneous solutions to eqns.~\eqref{eq:U1_W}
to~\eqref{eq:U1_visible}.

\subsection
[An \texorpdfstring{$E_7 \times U(1)$}{E7 x U1} Hidden Sector]
{An \texorpdfstring{\boldmath$E_7 \times U(1)$}{E7 x U1} Hidden Sector}

The first observation is that the system of equations \eqref{eq:U1_FI} to
\eqref{eq:U1_visible} is homogeneous with respect to the rescaling\footnote{Note
  that the coupling constants appear only in the combination
  $\frac{g_{s}^{2}l_{s}^{4}}{v^{2/3}}$, which is a positive number.}
\begin{equation}
  \left(
    a^1,~ 
    a^2,~
    a^3,~
    \frac{g_{s}^{2}l_{s}^{4}}{v^{2/3}}
  \right)
  \mapsto
  \left(
    \mu a^1,~ 
    \mu a^2,~
    \mu a^3,~
    \mu^2 \frac{g_{s}^{2}l_{s}^{4}}{v^{2/3}}
  \right)
  ,\quad
  \mu > 0
  .
  \label{eq:homogeneous}
\end{equation}
Therefore, one can absorb the coupling constants into the K\"ahler
moduli $a^i$.  In other words, for a single $U(1)$ the $FI=0$ equation
\eqref{eq:U1_FI} fixes one K\"ahler modulus to a certain numerical
value measured in multiples of
$\big(\frac{g_{s}^{2}l_{s}^{4}}{v^{2/3}}\big)^{1/2}$. This is
tantamount to setting $\frac{g_{s}^{2}l_{s}^{4}}{v^{2/3}}$ to unity,
which we will do henceforth for simplicity.

Next, let us concentrate on the particular hidden line bundle
\begin{equation}
  V^{(2)} = L
  = \Osheaf_X(1,2,3)
  ,
  \label{add1}
\end{equation}
that is, 
\begin{equation}
  (\ell^1, \ell^2, \ell^3) = (1,2,3)
  .
\end{equation}
This choice cancels the anomaly with the effective five-brane curve
class
\begin{equation}
  W = (16,10,0) \ge 0 .
\end{equation}
Having fixed the line bundle, we proceed to solve the Fayet-Iliopoulos
equation \eqref{eq:U1_FI} for $a^3$, giving us
\begin{equation}
  a^3 = \frac{
    -2 (a^1)^2-(a^2)^2-24 a^1 a^2-108 \lambda^2-108 \lambda +117
  }{
    6 \left(2 a^1+a^2\right)
  }.
  \label{eq:a3FIconstraint}
\end{equation}
This can then be substituted into equations \eqref{eq:U1_gvis},
\eqref{eq:U1_ghid} and~\eqref{eq:U1_visible} to obtain a system of
polynomial inequalities in $a^1, a^2$ and $\lambda$.

\begin{figure}[htbp]
   \center{
     \raisebox{1cm}{(a)}
     \conditionallyincludegraphics[width=0.4\textwidth]{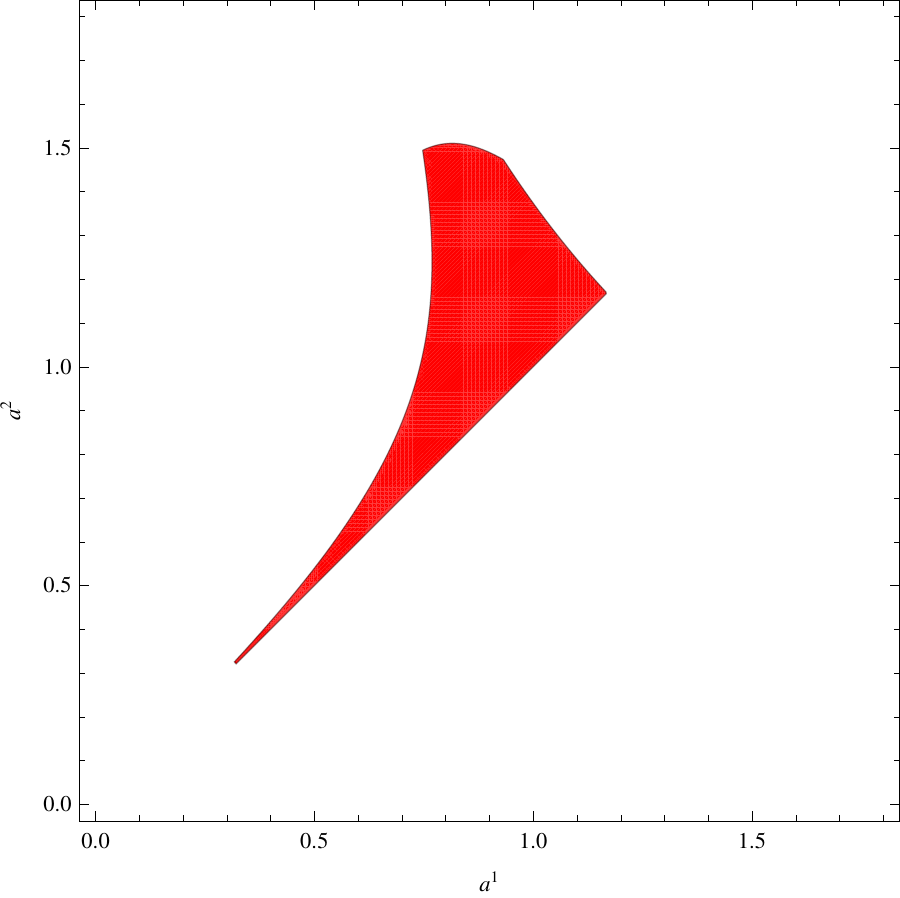}
     \hspace{1cm}
     \raisebox{1cm}{(b)}
     \conditionallyincludegraphics[width=0.4\textwidth]{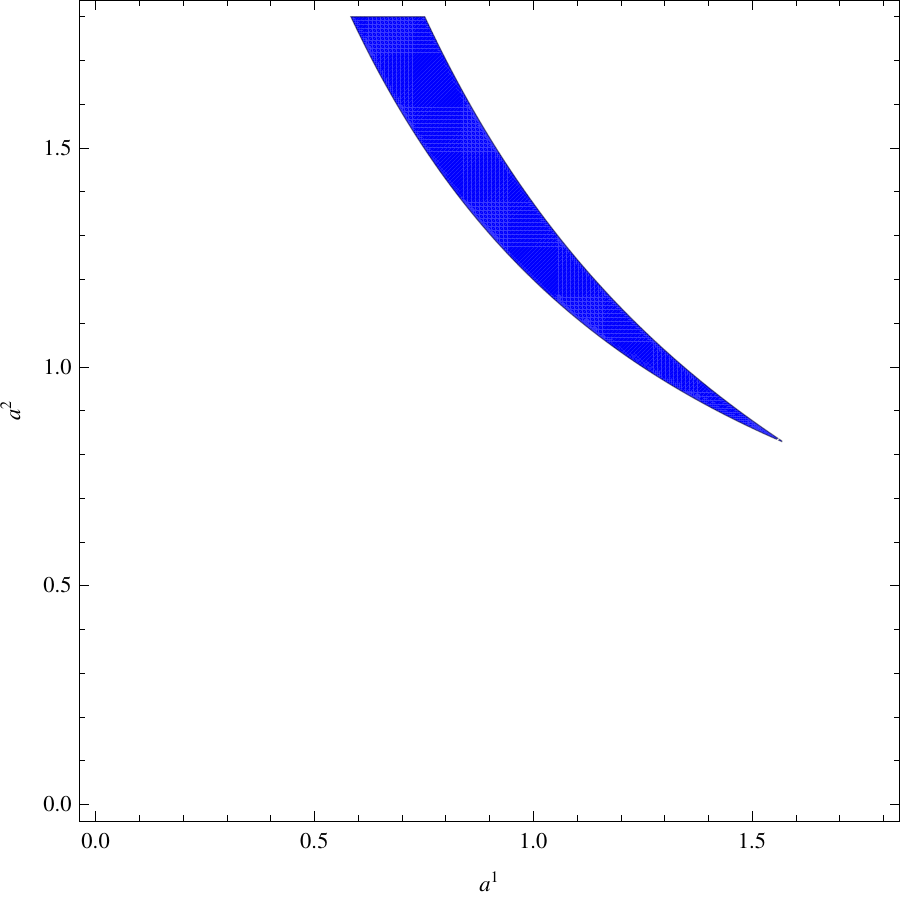}
   }
   \center{
     \raisebox{1cm}{(c)}
     \conditionallyincludegraphics[width=0.6\textwidth]{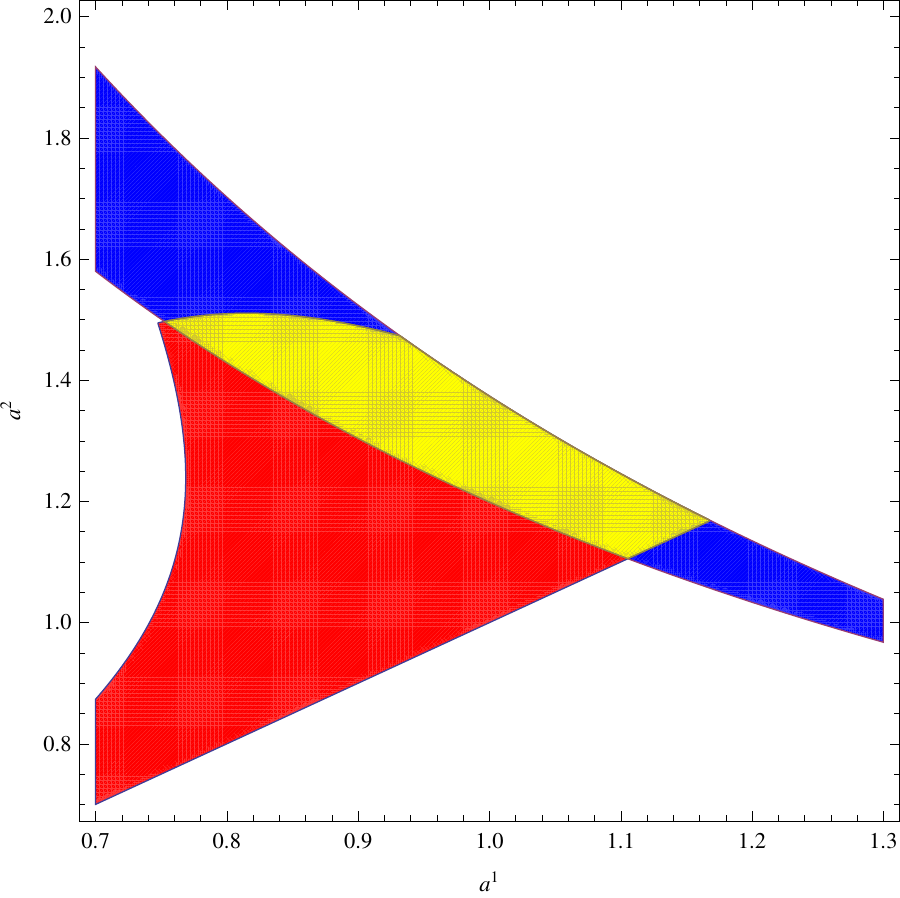}
   }
   \caption{The two-dimensional slice through the K\"ahler cone where
     the FI-term of the hidden line bundle $L = \Osheaf_X(1,2,3)$ with
     five-brane position $\lambda = 0.496$ vanishes. The slice is
     parametrized by $(a^1, a^2)$ with $a^3$ given by
     \eqref{eq:a3FIconstraint}. In red, the visible sector stability
     condition, see sub-figures a) and c). In blue, the region where
     the both the visible and hidden sector gauge couplings are
     positive, see sub-figures b) and c). Their intersection is drawn
     in yellow, see sub-figure c).}
   \label{fig:linesol}
\end{figure}
\begin{figure}[p]
  \setlength{\captionmargin}{0cm}
  \center{
    \vspace{-1cm}
    \conditionallyincludegraphics[width=0.9\textwidth,height=9cm]{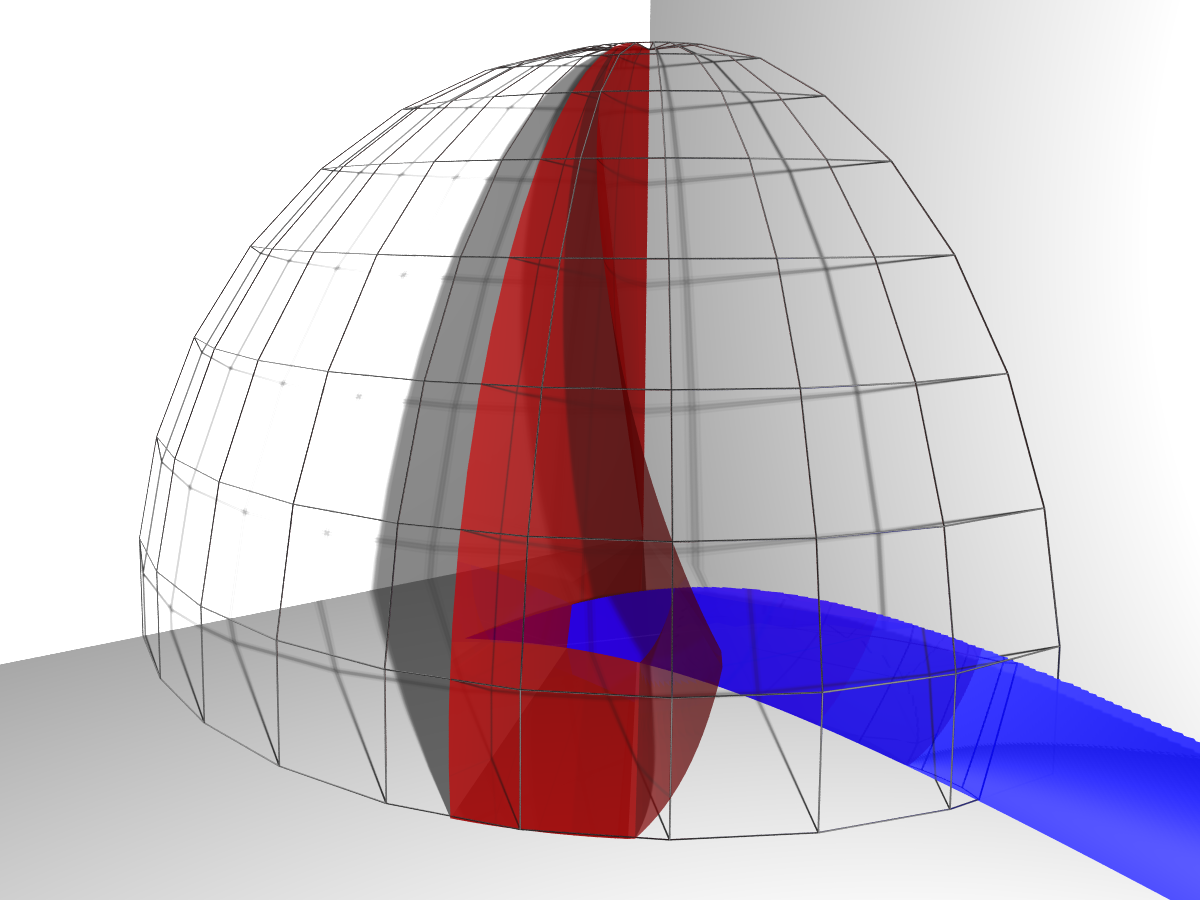}
    \conditionallyincludegraphics[width=0.9\textwidth,height=9cm]{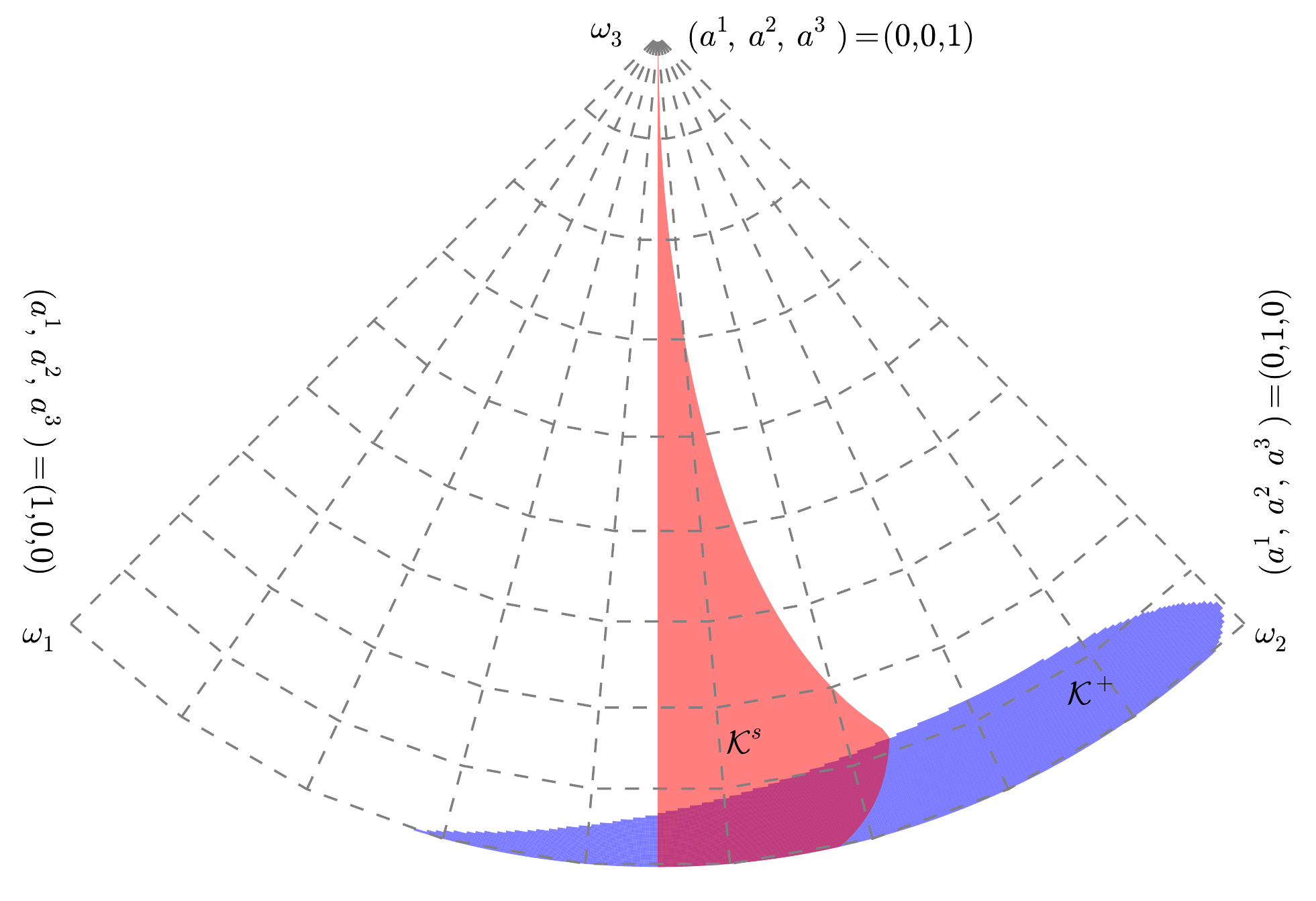}
  }
  \caption{The K\"ahler cone, in 3 dimensions (top) and the projection
    in radial directions (bottom). The blue region $\Kcone^+$ is our
    hidden sector solution for $L = \Osheaf_X(1,2,3)$ at $\lambda =
    0.496$. It shows the K\"ahler moduli $\omega =
    a^1\omega_1+a^2\omega_2+a^3\omega_3$ simultaneously satisfying the
    $FI=0$ condition and the positivity of the visible and hidden sector
    gauge couplings. The red region $\Kcone^s$ is the stability region
    of the visible sector bundle from \autoref{fig:starmap}. The
    intersection $\Kcone^s\cap \Kcone^+$ is where all physical
    constraints are satisfied.}
  \label{fig:hidden3d}
\end{figure}

Finally, one can scan through the range $-\frac12 < \lambda < \frac12$
and plot the region of validity in the $a^1$-$a^2$ plane to find
solutions. For example, with the numerical value of $\lambda = 0.496$,
which is close to the hidden wall, we do indeed find solutions which
satisfy all of the constraints. The region where all physical
constraints equations \eqref{eq:U1_gvis} to \eqref{eq:U1_visible} are
satisfied simultaneously is shown in yellow in \autoref{fig:linesol}.
For clarity, and to make contact with the visible sector stability
region plotted in \autoref{fig:starmap}, we present the same data in
\autoref{fig:hidden3d} as a subset of the K\"ahler cone---that is, the
positive octant $a^1$, $a^2$, $a^3> 0$. This gives a multi-dimensional
visualization of the facts that:
\begin{itemize}
\item The visible sector stability region (red) is a three-dimensional
  sub-cone of the K\"ahler cone.
\item The FI-term stabilizes one particular combination of angular and
  radial K\"ahler moduli, which is why the hidden sector $U(1)$
  constrains us to a two-dimensional surface (blue) in the K\"ahler
  cone.
\end{itemize}

\subsection{Hidden Sector Matter Spectrum}

We can now compute the low energy $U(1) \times E_7$ particle spectrum from the
cohomology of the line bundle $L$.  From the breaking pattern in
\eqref{73} and \eqref{74}, one can read off the representation
content \eqref{75}. The spectrum is then determined by the cohomology groups of
the corresponding tensor power of the line bundle. 
These are tabulated in the middle column of
\autoref{tab:E7U1spectrum}. They are valid for any line bundle $L$ appropriately embedded in the $SU(2)\subset E_{8}$.
\begin{table}[htbp]
  \centering
  \renewcommand{\arraystretch}{1.3}
  \begin{tabular}{@{$\quad$}c@{$\quad$}|@{$\quad$}c|@{$\qquad$}c@{$\quad$}}
    $U(1) \times E_7$ & 
    \multirow{1}{*}{Cohomology} & 
    Index ${\chi}$\\
    \hline
    $(0,\Rep{133})$ & 
    $H^*(X,{ \cal{O}}_{X})$ &
    ~~~$0$ \\
    $(1,\Rep{56})$  & 
    $H^*(X, L)$ &
   ~~ $8$ \\    
    $(-1,\Rep{56})$ & 
    $H^*(X, L^*)$ &
    $-8$ \\
    $(2,\Rep{1})$   & 
    $H^*(X, L^2)$ &
    ~~$58$ \\
    $(-2,\Rep{1})$  & 
    $H^*(X, L^{*2})$ & 
    $-58$ \\
    $(0,\Rep{1})$   & 
    $H^*(X, {\cal{O}}_{X})$ &
    ~~$0$ \\
  \end{tabular}
  \caption{The chiral spectrum for the $E_8\to U(1)\times E_7$ breaking
    pattern with a line bundle $L$. The index  $\chi$ counts the number
    of left-chiral minus the number of right-chiral fermionic zero-modes with the given gauge charge.}
\label{tab:E7U1spectrum}
\end{table}
In determining the bundles associated with a given representation, we used the fact that the line bundle $L$ induces a rank two bundle $L \oplus L^{*}$ with structure group $S(U(1) \times U(1)) \subset SU(2)$. Recalling that $L \otimes L^{*}={\cal{O}}_{X}$, it follows that
\begin{equation}
{\rm End} \big(L \oplus L^{*} \big)=(L \oplus L^{*}) \otimes (L \oplus L^{*})=(L^{2} \oplus L^{*2} \oplus {\cal{O}}_{X}) \oplus {\cal{O}}_{X} \ .
\label{AAA}
\end{equation}
Therefore 
\begin{equation}
H^{*}({\rm End} (L \oplus L^{*}))=\big(H^{*}(L^{2}) \oplus H^{*}(L^{*2}) \oplus H^{*}({\cal{O}}_{X})\big) \oplus H^{*}({\cal{O}}_{X}) ,
\label{BBB}
\end{equation}
corresponding to the $(2,\Rep{1}), (-2,\Rep{1}), (0,\Rep{1})$ and $(0,\Rep{133})$ representations respectively.

For each representation, the number of chiral fermionic zero-modes of
the Dirac operator is determined from the corresponding index
$\chi$. For an arbitrary vector bundle $V$ on a Calabi-Yau threefold,
the Atiyah-Singer index theorem tells us that
\begin{eqnarray}
 &&  \chi(V) = 
   \sum_{i=0}^3 
   (-1)^i h^i (X, V) = \int_X \ch(V)
   \wedge \td(TX) \nonumber \\
   &&~\quad \quad  =  \int_X 
    \left(
      \frac{1}{12} c_1(V) \wedge c_2(TX) + \ch_3(V)
    \right).
\end{eqnarray}
In the case of {\it any} single line bundle of the form
$V=\Osheaf_X(\ell^1,\ell^2,\ell^3)$, this simplifies to
\begin{equation}
  \chi( \Osheaf_X(\ell^1,\ell^2,\ell^3)) =  \frac13 \ell^1 + \frac13 \ell^2 +
    \frac16 \sum_{i,j,k=1}^3  d_{ijk}  \ell^i \ell^j \ell^k .
\label{ASI}
\end{equation}
Using this, the index of any of the tensor powers of $L$ appearing in the middle column of \autoref{tab:E7U1spectrum} is easily computed. 

Before restricting to the specific example specified in \eqref{add1}, let us present some important generic results. To begin with, for any  bundle $V$ 
\begin{equation}
\chi(V)=-\chi(V^{*}) \ .
\label{CCC}
\end{equation}
Therefore, for $V$ carrying a real representation of its structure group, that is, $V=V^{*}$, it follows that $\chi(V)=0$. For example
\begin{equation}
\chi({\cal{O}}_{X})=0 \ , 
\label{DDD}
\end{equation}
as indicated in the first and last entries of the third column of \autoref{tab:E7U1spectrum}. Hence, any possible fermionic  zero-modes associated with the representations $(0,\Rep{133})$ and $(0,\Rep{1})$ in the decomposition \eqref{75} must be non-chiral. In fact, one can determine the exact number of conjugate pairs of fermions there are in any such representation. Recall that the trivial line bundle $\Osheaf_X$ on a Calabi-Yau
threefold $X$ has cohomology groups of dimension 
\begin{equation}
h^0(X,\Osheaf_X) =
h^3(X,\Osheaf_X) = 1
\label{burtA}
\end{equation}
and $0$ otherwise. Note that this is consistent with a vanishing Atiyah-Singer index since $\chi=h^{0}-h^{1}+h^{2}-h^{3}$. It follows that for the trivial bundle ${\cal{O}}_{X}$, there is exactly one left-chiral fermion zero-mode specificed by $h^{0}=1$ and one conjugate right-chiral fermion zero-mode specified by $h^{3}=1$. These are the conjugate fermion gauginos in a vector supermultiplet. Specifically, our effective theory has one vector multiplet in the $\Rep{133}$ adjoint representation of $E_{7}$ and one vector multiplet in the adjoint $\Rep{1}$ representation of $U(1)$. 

A second generic result is that because of relation \eqref{CCC}, one needs to compute the index of only one bundle in each conjugate pair appearing in \autoref{tab:E7U1spectrum}. We will choose $L$ and $L^{2}$, corresponding to the representations $(1,\Rep{56})$ and $(2,\Rep{1})$ respectively. Consider, for example, the bundle $L$. Then $\chi (L)=n_{L}-n_{R}$ gives the number of chiral families of fermionic zero-modes transforming in the $(1,\Rep{56})$ representation. In principle, there can also be some number of vector-like pairs of such zero-modes. However, these pairs are naturally massive and, hence, we will ignore them. In fact, within the context of an \emph{ample} line bundle, such as \eqref{add1}, one can go further and actually compute the number of positive and negative chirality modes. To see this, note that by definition an ample line bundle has only positive entries $\ell^{1},\ell^{2},\ell^{3}$.
An immediate consequence is that
\begin{equation}
  \label{kodaira}
  h^i(X, L) = h^i(X,  L^{2}) = 0 , \quad i \not= 0
\end{equation}
by the Kodaira vanishing theorem. Therefore, only $h^0(X,L)$ and
$h^0(X,L^2)$ can be non-zero. Recalling that $\chi=h^{0}-h^1+h^2-h^3$, it follows from \eqref{kodaira} that for each of $L$ and $L^2$ there is exactly $\chi=h^{0}$ left-chiral fermion zero-modes and no right-chiral zero-modes. That is, there are no vector-like pairs. Note that this phenomenon of a non-vanishing index arising from $h^{0}$ (and, generically, $h^{3}$)--as opposed to $h^{1},h^{2}$--is due to the fact that $c_{1}(L) \neq 0$. 
That is, the hidden sector bundle is chosen not to be supersymmetric classically. $N=1$ supersymmetry is only restored by the non-vanishing one-loop corrections to the Fayet-Iliopoulos term rendering the D-term zero. 
Phrased another way, for stable $SU(n)$ bundles $V$ one has $c_1(V)=0$ and $h^0=h^3=0$. Hence, the chiral spectrum is coming from $h^1$ and $h^2$ only.
However, our bundles do not have vanishing first Chern class. It follows that $h^0$ and, in general, $h^3$ need not vanish. In principle, therefore, all 4 cohomology groups can contribute to the index.

To proceed, one must specify the line bundle $L$. Choose this to be \eqref{add1}. 
Using expression \eqref{ASI} for the Atiyah-Singer index, we find 
\begin{equation}
\chi(X,L) = 8~~ \text{for}~~ L=\Osheaf_X(1,2,3) 
\label{burtC}
\end{equation}
and, similarly, that
\begin{equation}
\chi(X,L^{2}) = 58 ~~\text{for}~~ L^2 = \Osheaf_X(2,4,6) ,
\label{burtD}
\end{equation}
as indicated in the third column of \autoref{tab:E7U1spectrum}. That is, the effective $U(1) \times E_{7}$ theory has 8 chiral supermultiplets transforming as $(1,\Rep{56})$ and 58 chiral supermultiplets transforming as $(2,\Rep{1})$.
 In
summary, the complete $U(1)\times E_7$ hidden sector spectrum
of our model is
\begin{equation}
  1 \times (0,\Rep{133})
  +
  8 \times (1,\Rep{56})
  +
  58 \times (2,\Rep{1})
  + 
  1 \times (0,\Rep{1}) .
\end{equation}

\subsection{Constraints for Two Line Bundles}

In the previous section, we discussed the case of the hidden
sector being a single line bundle and presented a detailed solution.
One can easily move on to bundles of higher rank and show that indeed
there is a plenitude of solutions. This is clearly desirable for model
building since each independent $U(1)$ imposes one Fayet-Iliopoulos vanishing equation
and, therefore, stabilizes one K\"ahler modulus. In this section, we will consider the next simplest hidden sector consisting of the direct sum of two line bundles
\begin{equation}
V^{(2)}=L_{1} \oplus L_{2}, \qquad L_{1}={\cal{O}}_{X}(\ell_{1}^{1},\ell_{1}^{2},\ell_{1}^{3}),~L_{2}={\cal{O}}_{X}(\ell_{2}^{1},\ell_{2}^{2},\ell_{2}^{3})
\label{burtE}
\end{equation}
where 
\begin{equation}
\ell_{r}^{1}+\ell_{r}^{2}=0~{\rm mod}~3, \quad r=1,2 .
\label{burtF}
\end{equation}
Furthermore, $L_{1} \oplus L_{2}$ will be given the simplest simplest embedding into $E_8$, namely via
\begin{equation}
  U(1) \times U(1) \times SO(12)
  ~\subset~
  SU(2) \times SU(2) \times SO(12) 
  ~\subset~
  E_8
  .
\end{equation}
The branching rules easily follow from those of the $SU(2) \times
SO(12)$ maximal subgroup of $E_7$ in conjunction with \eqref{73}. In
particular, the adjoint representation of $E_8$ decomposes under $SU(2) \times SU(2) \times SO(12)$ as
\begin{equation}
  \begin{split}
    \Rep{248} \longrightarrow\;&
    (\Rep{1}, \Rep{3}, \Rep{1}) \oplus 
    (\Rep{3}, \Rep{1}, \Rep{1}) \oplus
    (\Rep{2}, \Rep{1}, \Rep{32}) \oplus
    (\Rep{1}, \Rep{2}, \barRep{32})
    \\&
    \oplus
    (\Rep{2}, \Rep{2}, \Rep{12})\oplus
    (\Rep{2}, \Rep{2}, \Rep{12}) \oplus 
    (\Rep{1}, \Rep{1}, \Rep{66}) .
  \end{split}
  \label{burtG}
\end{equation}

One now has to choose the embedding of the generator $Q_r$ of each
of the two $U(1)$ structure groups into the corresponding $SU(2)$. We again pick the simplest
embedding, identifying the structure group of $L_1$ with the center of
the first $SU(2)$, as in \eqref{74}, and similarly for $L_2$.
Consequently, under $U(1) \times U(1) \times SO(12)$ we have the
branching rule
\begin{equation}
  \begin{split}
    \Rep{248} \longrightarrow\;&
    (0, 2, \Rep{1}) \oplus 
    (2,0,\Rep{1}) \oplus 
    (0,-2,\Rep{1}) \oplus 
    (-2,0,\Rep{1}) \oplus
    2 \times (0,0,\Rep{1})
    \\&
    \oplus
    (1,0,\Rep{32}) \oplus (-1,0,\Rep{32}) \oplus
    (0,1,\barRep{32}) \oplus (0,-1,\barRep{32}) \oplus
    (1,1,\Rep{12}) 
    \\&
    \oplus  (1,-1,\Rep{12}) \oplus 
    (-1,1,\Rep{12}) \oplus
    (-1,-1,\Rep{12})
    \oplus 
    (0,0,\Rep{66}) .
  \end{split}
  \label{burtH}
\end{equation}
As before, one can read off from \eqref{26} the group-theoretic
embedding coefficients $a_1$ and $a_2$. They are given by 
\begin{equation}
a_1 = a_2 = 1 . 
\label{burtI}
\end{equation}
Again
assuming that we have only a single fivebrane, the constraints for the
case of the direct sum of two line bundles---analogous
to equations \eqref{eq:U1_W} to~\eqref{eq:U1_visible}---are
\begin{subequations}
  \begin{align}
    W_{i} = \big(\tfrac{4}{3},\tfrac{7}{3},-4\big)_{i} +
    \sum\limits_{r=1}^2 d_{ijk} \ell^j_r \ell^k_r 
    &\;\geq 0
    , \quad i=1,2,3 
    \label{eq:twolb_W}
    \\[1ex]
    d_{ijk}\ell^{i}_r a^{j}a^{k}- \frac{g_{s}^{2}l_{s}^{4}}{v^{2/3}}\Big( 
    d_{ijk}\ell^{i}_r \ell^{j}_r \ell^{k}_r + \ell^{i}_r (2,2,0)_i
    \qquad &\nonumber \\
    -\big(\tfrac{1}{2}+\lambda\big)^{2}
    \ell^{i}_r W_{i}\Big) 
    &\;= 0
    ,  \quad r = 1,2 
    \label{eq:twolb_FI} 
    \\[1ex]
    d_{ijk} a^i a^j a^k
    - 3  \frac{g_{s}^{2}l_{s}^{4}}{v^{2/3}} \Big(
    -\big(\tfrac83 a^1 + \tfrac53 a^2 + 4 a^3\big)
    \qquad &\nonumber \\
    + 2(a^1+a^2)
    -\big(\tfrac{1}{2}-\lambda\big)^{2} a^i W_i\Big) 
    &\;> 0 
    ,
    \label{eq:twolb_gvis}
    \\[1ex]
    d_{ijk} a^i a^j a^k- 3  \frac{g_{s}^{2}l_{s}^{4}}{v^{2/3}}
    \Big(
    d_{ijk}a^i \ell^j \ell^k
    + 2(a^1+a^2)
    \qquad &\nonumber \\
    -\big(\tfrac{1}{2}+\lambda\big)^{2} a^i W_i\Big) 
    &\;> 0 
    , 
    \label{eq:twolb_ghid}
  \end{align}
  \begin{multline}
    \VisibleStabilityCond
    \label{eq:twolb_visible}
  \end{multline}
\end{subequations}
We point out that only the first two sets of equations differ from the
single line bundle case in the previous section.  We must now find
simultaneous solutions to equations \eqref{eq:twolb_W}
to~\eqref{eq:twolb_visible}.

\subsection
[An \texorpdfstring{$SO(12)\times U(1)\times U(1)$}{SO(12)xU(1)xU(1)} Hidden Sector]
{An \texorpdfstring{\boldmath$SO(12)\times U(1)\times U(1)$}{SO(12)xU(1)xU(1)} Hidden Sector}

We will now focus on the specific direct sum of two line bundles given by
\begin{equation}
  V^{(2)} = L_1 \oplus L_2 = \Osheaf_X(2,1,1) \oplus \Osheaf_X(0,3,2) .
\end{equation}
In terms of our basis choice for the first Chern class,
see \eqref{eq:V2ell}, this is 
\begin{equation}
  (\ell^1_1, \ell^2_1, \ell^3_1) = (2,1,1) 
  , \quad
  (\ell^1_2, \ell^2_2, \ell^3_2) = (0,3,2) 
  .
\end{equation}
For simplicity, we assume that there is only a single five-brane. In
order to cancel the heterotic anomaly for this choice of bundle and
embedding, it must then wrap the effective curve class
\begin{equation}
  W = (20, 9, 0) \ge 0
  .
\end{equation}

As in the previous section, we use the homogeneous rescaling
\eqref{eq:homogeneous} to set $\frac{g_{s}^{2}l_{s}^{4}}{v^{2/3}}
= 1$. We will also use the same five-brane position for convenience;
namely $\lambda = 0.496$. The only remaining parameters are now the
K\"ahler moduli $a^1$, $a^2$ and $a^3$, subject to two FI-term constraints and a
number of inequalities. We first consider the two FI-term constraints
\eqref{eq:twolb_FI}, which are two quadratic equations that
stabilize two of the three K\"ahler moduli. The standard strategy to
parametrize the solution set is to pick one variable and then compute
a lexicographic Gr\"obner basis with the chosen variable last. If we
pick $a^3$ as the parameter, then the result is that $a^2$ is a real
solution of the quartic equation
\begin{figure}[p]
  \setlength{\captionmargin}{0cm}
  \center{
    \vspace{-1cm}
    \conditionallyincludegraphics[width=0.9\textwidth,height=9cm]{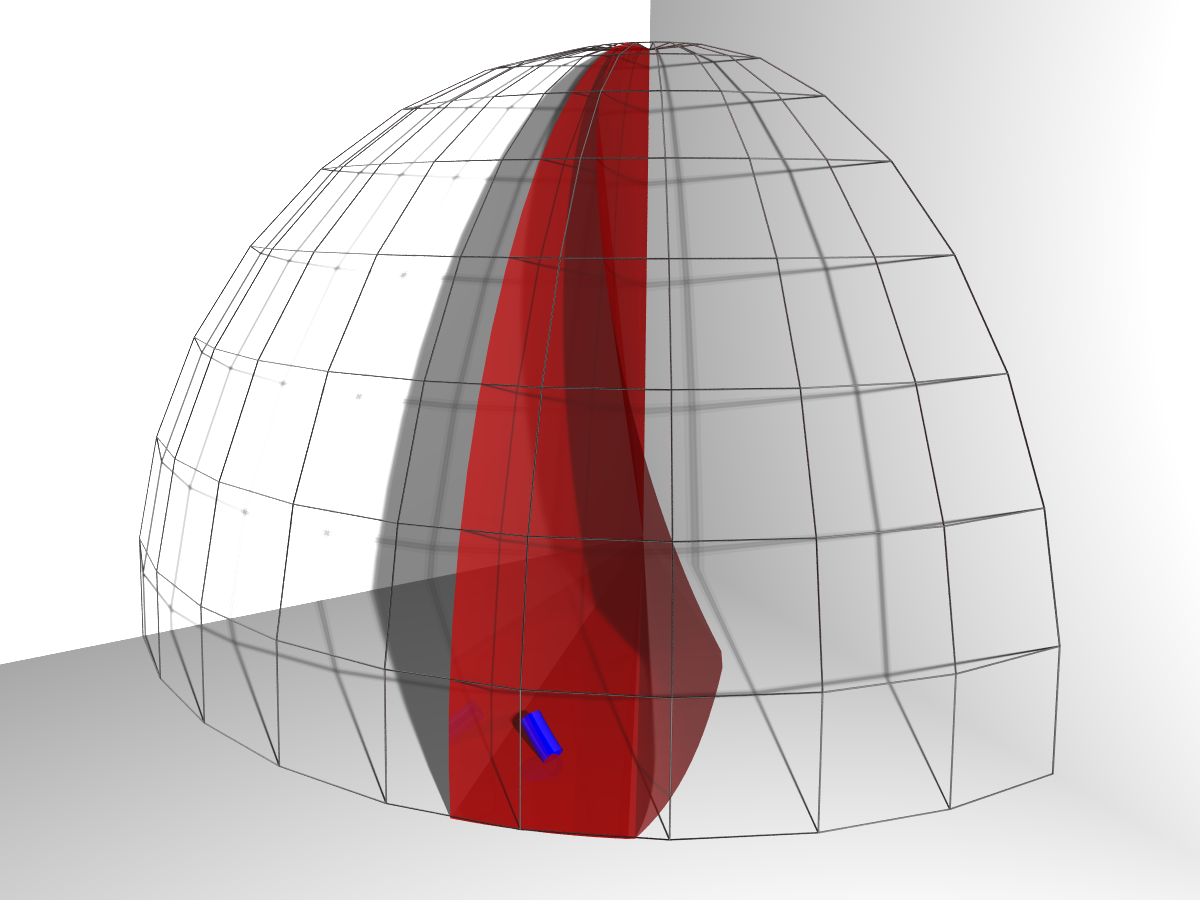}
    \conditionallyincludegraphics[width=0.9\textwidth,height=9cm]{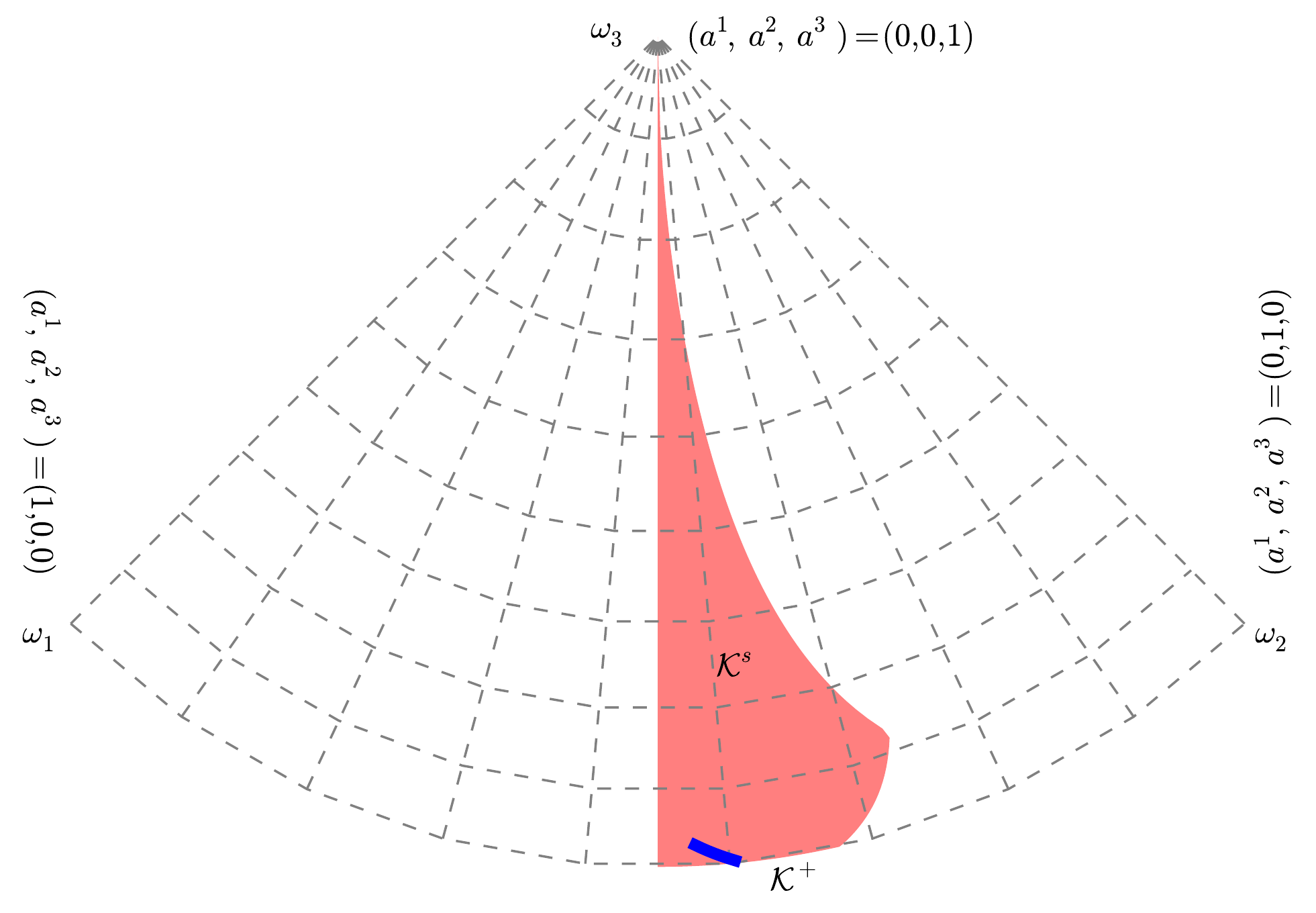}
  }
  \caption{The K\"ahler cone, in 3 dimensions (top) and the projection in radial
    directions (bottom). The 1-dimensional blue region $\Kcone^+$ is our hidden
    sector solution for $L_1\oplus L_2 = \Osheaf_X(2,1,1) \oplus
    \Osheaf_X(0,3,2)$ at $\lambda = 0.496$. It shows the K\"ahler
    moduli $\omega = a^1\omega_1+a^2\omega_2+a^3\omega_3$
    simultaneously satisfying the two independent $FI=0$ conditions for the two
    $U(1)$ factors, as well as the positivity of the isible and hidden sector
    gauge couplings. The red region $\Kcone^s$ is the stability region
    of the visible sector bundle from \autoref{fig:starmap}. The
    intersection $\Kcone^s\cap \Kcone^+$ is where all physical
    constraints are satisfied.}
  \label{fig:twolb3d}
\end{figure}
\begin{subequations}
  \begin{equation}
    (a^2)^4+6.7058 a^3 (a^2)^3+4.2352 (a^3)^2 (a^2)^2-0.3955 
    (a^2)^2-1.5808 a^3 a^2-1.1801 = 0
    \label{eq:a2}
  \end{equation}
  and, finally,
  \begin{equation}
    a^1 = 
    1.2651 (a^2)^3 + 8.4839 a^3 (a^2)^2 + 5.3582 (a^3)^2 a^2
    - 0.83377 a^2 - 4 a^3 
    .
    \label{eq:a1}
  \end{equation}
\end{subequations}
It remains to impose all inequalities; namely, the positivity of all
K\"ahler moduli, the gauge couplings equations \eqref{eq:twolb_gvis}
and \eqref{eq:twolb_ghid}, as well as the visible sector stability
conditions \eqref{eq:twolb_visible}. The numerical result is that the
free K\"ahler modulus has to lie in the interval
\begin{equation}
  0 < a^3 < 0.0701743
  .
\end{equation}
It then follows that the unique positive root $a^2$ of \eqref{eq:a2} and $a^1$
determined by \eqref{eq:a1} satisfy all physical constraints. For
example, if we pick $a^3 = 0.06$ then the remaining K\"ahler moduli and
gauge couplings are
\begin{equation}
  \begin{split}
    (a^1,a^2,a^3) = &\, (0.95, 1.06, 0.06) 
    , \\
    (
      \frac{4\pi}{g^{(1)2}}, \frac{4\pi}{g^{(2)2}} 
    ) =&\, (0.65, 0.02) 
    .
  \end{split}
\end{equation}
The entire one-dimensional solution set is plotted in
\autoref{fig:twolb3d}. 

Finally, as in the one line bundle case, the particle spectrum of this low energy $SO(12) \times U(1) \times U(1)$ hidden sector can be computed from the cohomology  of the various tensor products of $L_{1}$ and $L_{2}$. Since this is similar to the discussion in Subsection 4.3, but rendered more complicated by the presence of two line bundles, we will not discuss the results here.

\section*{Acknowledgments}

We would like to thank Lara Anderson, Ralph Blumenhagen, James Gray,
and Andre Lukas for useful discussions. Volker Braun is supported by
the Dublin Institute for Advanced Studies and expresses his thanks to
the University of Pennsylvania for its hospitality while some this
work was being carried out. Burt Ovrut is supported in part by the DOE
under contract No. DE-AC02-76-ER-03071 and the NSF under grant
No. 1001296. Yang-Hui He would like to thank the Science and
Technology Facilities Council, UK, for an Advanced Fellowship and
grant ST/J00037X/1, the Chinese Ministry of Education, for a
Chang-Jiang Chair Professorship at NanKai University, the
U.S. National Science Foundation for grant CCF-1048082, as well as
City University, London and Merton College, Oxford, for their enduring
support.

\begin{appendices}
\section{Line Bundles and 
  \texorpdfstring{\boldmath$\mathrm{E}_8$}{E8}}
\label{sec:linebundles}

\subsection{Induced Bundles}
\label{sec:induced}

The usual approach of constructing $E_8$ bundles for heterotic strings
is to first construct a $G$-bundle for a smaller group $G$ and then
use a map (group homomorphism) $\psi:G\to E_8$ to build an induced
$E_8$ bundle. Really, this is just the composition of the $G$-bundle
presentation $[X,BG]$ with $B\psi:BG\to BE_8$. Explicitly, we can
think of the $G$-bundle as a collection of transition functions
$\varphi_{\alpha\beta}: U_{\alpha\beta}\to G$ on overlaps
$U_{\alpha\beta}=U_\alpha\cap U_\beta$. The transition functions of
the induced $E_8$ bundle are then simply given by the composition
$\tilde{\varphi}_{\alpha\beta} = \psi \circ \varphi_{\alpha\beta}$. A
popular choice for $G$ is $SU(n)$, equivalent to a rank-$n$ vector
bundle of vanishing first Chern class. This is usually combined with a
group homomorphism $SU(n) \subset SU(9) \to SU(9)/\Z_3 \subset E_8$
that factors through the $SU(9)/\Z_3$ subgroup,\footnote{Because it
  will be important in this appendix, we will break with the physics
  tradition and be careful about discrete quotients of groups. A good
  reference for the relevant group theory is~\cite{Distler:2007av}.}
corresponding to removal of a node in $\tilde{E}_8$ affine Dynkin
diagram.

However, there is no need to pick a special unitary group and, for the
purposes of this section, we are considering $G=U(1)$. As a first
example, the easiest group homomorphism to $E_8$ can be obtained from
embedding $U(1)$ as a maximal torus in $SU(2)$ and then embedding it
further in $SU(9) \to E_8$. Up to a choice of coordinates, this is the
homomorphism
\begin{equation}
  \label{eq:U1SU2}
  U(1) \to SU(2), \quad
  e^{i \phi} \mapsto 
  \begin{pmatrix}
    \exp(i\phi) & 0 \\ 0 & \exp(-i\phi)
  \end{pmatrix}
\end{equation}
or 
\begin{math}
  1 \mapsto
  \left(\begin{smallmatrix}
    1 & 0 \\ 0 & -1
  \end{smallmatrix}\right)
\end{math}
in the Lie algebra. This construction cannot yield the most general
$E_8$ bundle, but rather only one whose structure group can be reduced
to $SU(9)/\Z_3 \subset E_8$. This is of course desirable for
phenomenological applications, so that not all of the $E_8$ gauge
group is broken. In this example, the commutant of $SU(2)$ in $E(8)$
is $E_7/\Z_2$ and the embedded $U(1)\subset SU(2)$ commutes with it
self. Hence, the overall commutant and unbroken gauge group of
$U(1)\subset E_8$ is $(E_7\times U(1))/\Z_2$.

\subsection{Consistency Conditions}
\label{sec:consistency}

It is occasionally claimed that a $U(n)$ bundle, for example, a line
bundle, has to satisfy extra divisibility conditions in order to
define an induced $E_8$ bundle. However, this is not true and any
bundle together with a $U(n)\to E_8$ homomorphism is admissible. In
particular, the first Chern class of the line bundle need \emph{not}
be even.\footnote{That is, divisible by $2$.} An $E_8$ bundle is
automatically spin since $E_8$ is simply connected. While it is true
that a line bundle with an odd first Chern class is not spin, any
induced $E_8$ bundle is well-defined and admits adjoint
spinors.\footnote{We remark that this is different in the
  $\Spin(32)/\Z_2$ heterotic string if one uses a line bundle and
  $U(1)\to SO(32)$ homomorphism. In this case, the line bundle does
  have to satisfy additional constraints for the induced bundle to
  lift to a $\Spin(32)/\Z_2$ bundle.} Since the $U(n)$ bundle is
purely auxiliary in this construction, there is no need for it to be
spin. This is related to the fact that the contribution to the
heterotic anomaly from a line bundle is always an even multiple of its
second Chern character, as we will discuss in \autoref{sec:U1chern}.

There is a related, but different, context where an even first Chern
class does play a role~\cite{Distler:1987ee, Freed:1986zx}. In an
effort to clear up any confusion, let us review it in the remainder of
this subsection. First, recall the usual conformal field theory
construction of the $E_8\times E_8$ heterotic string. There are 36
real left-moving fermions, $16$ for each $E_8$. The GSO projection
acts as a minus sign, so the $16$ fermions transform under
$\Spin(16)/\Z_2$. Only this group, and not the whole $\Spin(16)$, is a
subgroup of $E_8$. Since it is still difficult to construct the most
general $\Spin(16)/\Z_2$ bundle, it is tempting to combine the $16$
real spinors into $8$ complex ones and construct a $U(8)$ gauge bundle
for them. The trouble is that this constructs a $U(8) \subset SO(16)$
gauge bundle which need not be a $\Spin(16)/\Z_2$ bundle. A sufficient
condition to avoid this problem is if the $U(8)$ bundle is spin, that
is, has even first Chern class. In that case it lifts to a $\Spin(16)$
bundle, which we can divide to obtain a $\Spin(16)/\Z_2$ bundle. The
key difference is that this does not use a $U(8)\to
\Spin(16)/\Z_2\subset E_8$ homomorphism to construct an induced $E_8$
bundle, but rather relies on a lifting that might not exist.

\subsection{Chern Classes and the Anomaly}
\label{sec:U1chern}

The heterotic anomaly cancellation condition for two $E_8$-bundles
$\Vvis$, $\Vhid$ is
\begin{equation}
  -c_2(\Vvis)   -c_2(\Vhid)  +  c_2(TX) = W 
  \quad
  \in H^4(X,\Z)
  ,
\end{equation}
where $W$ is the the class of the five-brane(s). Here, $c_2$ of a
$E_8$ bundle means its degree-$4$ characteristic class. Completely
analogous to the usual Chern classes, it is unambiguously defined in
integral cohomology as the pull-back of the generator $c_2\in
H^4(BE_8,\Z) \simeq \Z$ by the map $[X, BE_8]$ defining the
bundle. Fortunately, we never have to actually evaluate the
homotopy-theoretic definition. For an $E_8$ bundle whose structure
group reduces to the $SU(9)/\Z_3$ subgroup, which is the case we are
interested in for phenomenolgical reasons, the degree-4 characteristic
class of the $E_8$ bundle coincides with the usual second Chern class
of the $SU(9)$ bundle. Hence, the contribution to the anomaly of such
an induced bundle $V_\rho$ defined by a $U(n)$ bundle $V$ and
homeomorphism $\rho: U(n)\to SU(9)$ can be computed in terms of the
Chern classes of $V$ and the group theory of $\rho$.

Let us consider the case where $\Vvis = L_\rho$ is induced from
a line bundle $L$ and $\rho:U(1)\to SU(9)$. On general grounds,
the anomaly cancellation condition then must be of the form
\begin{equation}
  a_\rho\; c_1(L)^2 - c_2(\Vhid) + c_2(TX) = W 
  \quad
  \in H^4(X,\Z)
\end{equation}
since the only available characteristic class is $c_1(L)$ with
some group-theoretic numerical coefficient $a_\rho$. In de Rham
cohomology the visible sector contribution is represented by
\begin{equation}
  a_\rho \; c_1(L)^2 = 
  2 a_\rho\; \ch_2(L) = 
  \frac{1}{16 \pi^2} \tr_{\rho} F \wedge F
\end{equation}
It is suggestive, but wrong, that the coefficient $a_\rho$ should be
half-integral such that the contribution of the line bundle to the
anomaly is always an integer multiple of its second Chern
character. In fact, the coefficient $a_\rho$ is always integer which
is twice what one would naively expect.\footnote{For discrete Wilson
  lines, this observation was already made in the footnote on Page 88
  of~\cite{Witten:1985mj}.} This is also required for the anomaly
contribution $a_\rho c_1(L)^2$ to define an integral cohomology
class. As the simplest example, let us return to the $\rho: U(1)\to
SU(2) \subset SU(9)$ embedding from \eqref{eq:U1SU2}. The induced
$SU(9)$ bundle is
\begin{equation}
  L_\rho = L \oplus L^{-1} \oplus \mathbf{1}_7,
\end{equation}
so its second Chern class is 
\begin{math}
  - c_2(L_\rho) = 
  c_1^2(L),
\end{math}
that is, $a_\rho = 1$.

\subsection{Example}


As a more complicated example, we now consider a combination of line
bundle and non-Abelian bundle. Let us start with a $SU(3)$ bundle $V$
and a line bundle $L$. We can use this data with different group
homomorphisms to construct the same $SU(9)$ (and, therefore, $E_8$)
bundle in two different ways:
\begin{enumerate}[(A)]
\item Use the group homomorphism
  \begin{equation}
    \begin{split}
      \rho_A: SU(3) \times U(1) &\to SU(4) \subset SU(9), 
      \\ 
      \big( g_{3\times 3}, e^{i\phi} \big)
      &\mapsto 
      \begin{pmatrix}
        e^{i\phi} g_{3\times 3} & 0 & 0 \\ 
        0 & e^{-3i\phi} & 0 \\
        0 & 0 & \mathbf{1}_{5\times 5}
      \end{pmatrix}
    \end{split}
  \end{equation}
\item The direct sum $(V\otimes L) \oplus L^{-3}$ is a
  rank-$4$ bundle with vanishing first Chern class. We combine it with
  the trivial embedding $\rho_B:SU(4)\subset SU(9)$.
\end{enumerate}
Both constructions yield the same induced $SU(9)$-bundle, namely
\begin{equation}
  E =
  (V\otimes L) \oplus L^{-3} \oplus \mathbf{1}_5
  = 
  (V \oplus L)_{\rho_A}
  =
  \big( (V\otimes L) \oplus L^{-3} \big)_{\rho_B}
  .
\end{equation}
Its contribution to the heterotic anomaly is
\begin{math}
  -c_2(E) = -c_2(V) + 3 c_1(L)^2
  .
\end{math}
In other words, the group-theoretic coefficient $a_\rho=3$. It is
again an integral class, as it must be.

\end{appendices}


\renewcommand{\refname}{Bibliography}
\addcontentsline{toc}{section}{Bibliography} 
\bibliographystyle{utcaps} 
\bibliography{WeakHetPaperFinal3.bib}

\providecommand{\href}[2]{#2}\begingroup\raggedright\begin{thebibliography}{10}

\bibitem{Gross:1985fr}
D.~J. Gross, J.~A. Harvey, E.~J. Martinec, and R.~Rohm, ``Heterotic String
  Theory. 1. The Free Heterotic String,'' {\em Nucl. Phys.} {\bf B256} (1985)
253.

\bibitem{Gross:1985rr}
D.~J. Gross, J.~A. Harvey, E.~J. Martinec, and R.~Rohm, ``Heterotic String
  Theory. 2. The Interacting Heterotic String,'' {\em Nucl. Phys.} {\bf B267}
  (1986)
75.

\bibitem{Horava:1995qa}
P.~Horava and E.~Witten, ``Heterotic and type {I} string dynamics from eleven
  dimensions,'' {\em Nucl. Phys.} {\bf B460} (1996) 506--524,
\href{http://arXiv.org/abs/hep-th/9510209}{{\tt hep-th/9510209}}.

\bibitem{Horava:1996ma}
P.~Horava and E.~Witten, ``{Eleven-dimensional supergravity on a manifold with
  boundary},'' {\em Nucl.Phys.} {\bf B475} (1996) 94--114,
\href{http://arXiv.org/abs/hep-th/9603142}{{\tt hep-th/9603142}}.

\bibitem{Lukas:1997fg}
A.~Lukas, B.~A. Ovrut, and D.~Waldram, ``On the four-dimensional effective
  action of strongly coupled heterotic string theory,'' {\em Nucl. Phys.} {\bf
  B532} (1998) 43--82,
\href{http://arXiv.org/abs/hep-th/9710208}{{\tt hep-th/9710208}}.

\bibitem{Lukas:1998ew}
A.~Lukas, B.~A. Ovrut, and D.~Waldram, ``The ten-dimensional effective action
  of strongly coupled heterotic string theory,'' {\em Nucl. Phys.} {\bf B540}
  (1999) 230--246,
\href{http://arXiv.org/abs/hep-th/9801087}{{\tt hep-th/9801087}}.

\bibitem{Lukas:1998hk}
A.~Lukas, B.~A. Ovrut, and D.~Waldram, ``{Nonstandard embedding and five-branes
  in heterotic M theory},'' {\em Phys.Rev.} {\bf D59} (1999) 106005,
\href{http://arXiv.org/abs/hep-th/9808101}{{\tt hep-th/9808101}}.

\bibitem{Lukas:1998tt}
A.~Lukas, B.~A. Ovrut, K.~Stelle, and D.~Waldram, ``{Heterotic M theory in
  five-dimensions},'' {\em Nucl.Phys.} {\bf B552} (1999) 246--290,
\href{http://arXiv.org/abs/hep-th/9806051}{{\tt hep-th/9806051}}.

\bibitem{Lukas:1998yy}
A.~Lukas, B.~A. Ovrut, K.~Stelle, and D.~Waldram, ``{The Universe as a domain
  wall},'' {\em Phys.Rev.} {\bf D59} (1999) 086001,
\href{http://arXiv.org/abs/hep-th/9803235}{{\tt hep-th/9803235}}.

\bibitem{MR86h:58038}
S.~K. Donaldson, ``Anti self-dual {Y}ang-{M}ills connections over complex
  algebraic surfaces and stable vector bundles,'' {\em Proc. London Math. Soc.
  (3)} {\bf 50} (1985), no.~1, 1--26.

\bibitem{MR88i:58154}
K.~Uhlenbeck and S.-T. Yau, ``On the existence of hermitian-{Y}ang-{M}ills
  connections in stable vector bundles,'' {\em Comm. Pure Appl. Math.} {\bf 39}
  (1986), no.~S, suppl., S257--S293. Frontiers of the mathematical sciences:
  1985 (New York, 1985).

\bibitem{Witten:1996mz}
E.~Witten, ``{Strong coupling expansion of Calabi-Yau compactification},'' {\em
  Nucl.Phys.} {\bf B471} (1996) 135--158,
\href{http://arXiv.org/abs/hep-th/9602070}{{\tt hep-th/9602070}}.

\bibitem{Greene:1986ar}
B.~R. Greene, K.~H. Kirklin, P.~J. Miron, and G.~G. Ross, ``A Superstring
  Inspired Standard Model,'' {\em Phys. Lett.} {\bf B180} (1986)
69.

\bibitem{Greene:1986bm}
B.~R. Greene, K.~H. Kirklin, P.~J. Miron, and G.~G. Ross, ``A Three Generation
  Superstring Model. 1. Compactification And Discrete Symmetries,'' {\em Nucl.
  Phys.} {\bf B278} (1986)
667.

\bibitem{Greene:1986jb}
B.~R. Greene, K.~H. Kirklin, P.~J. Miron, and G.~G. Ross, ``A Three Generation
  Superstring Model. 2. Symmetry Breaking And The Low-Energy Theory,'' {\em
  Nucl. Phys.} {\bf B292} (1987)
606.

\bibitem{Matsuoka:1986vg}
T.~Matsuoka and D.~Suematsu, ``Realistic Models From The {$E_8\times E_8'$}
  Superstring Theory,'' {\em Prog. Theor. Phys.} {\bf 76} (1986)
886.

\bibitem{Greene:1987xh}
B.~R. Greene, K.~H. Kirklin, P.~J. Miron, and G.~G. Ross, ``{$27^3$} {Yukawa}
  Couplings For a Three Generation Superstring Model,'' {\em Phys. Lett.} {\bf
  B192} (1987)
111.

\bibitem{Braun:2005nv}
V.~Braun, Y.-H. He, B.~A. Ovrut, and T.~Pantev, ``{The Exact MSSM spectrum from
  string theory},'' {\em JHEP} {\bf 0605} (2006) 043,
\href{http://arXiv.org/abs/hep-th/0512177}{{\tt hep-th/0512177}}.

\bibitem{Braun:2006ae}
V.~Braun, Y.-H. He, and B.~A. Ovrut, ``{Stability of the minimal heterotic
  standard model bundle},'' {\em JHEP} {\bf 0606} (2006) 032,
\href{http://arXiv.org/abs/hep-th/0602073}{{\tt hep-th/0602073}}.

\bibitem{Braun:2006me}
V.~Braun, Y.-H. He, and B.~A. Ovrut, ``{Yukawa couplings in heterotic standard
  models},'' {\em JHEP} {\bf 0604} (2006) 019,
\href{http://arXiv.org/abs/hep-th/0601204}{{\tt hep-th/0601204}}.

\bibitem{Gray:2007zza}
J.~Gray, A.~Lukas, and B.~Ovrut, ``{Perturbative anti-brane potentials in
  heterotic M-theory},'' {\em Phys.Rev.} {\bf D76} (2007) 066007,
\href{http://arXiv.org/abs/hep-th/0701025}{{\tt hep-th/0701025}}.

\bibitem{Gray:2007qy}
J.~Gray, A.~Lukas, and B.~Ovrut, ``{Flux, gaugino condensation and anti-branes
  in heterotic M-theory},'' {\em Phys.Rev.} {\bf D76} (2007) 126012,
\href{http://arXiv.org/abs/0709.2914}{{\tt 0709.2914}}.

\bibitem{Braun:2006th}
V.~Braun and B.~A. Ovrut, ``{Stabilizing moduli with a positive cosmological
  constant in heterotic M-theory},'' {\em JHEP} {\bf 0607} (2006) 035,
\href{http://arXiv.org/abs/hep-th/0603088}{{\tt hep-th/0603088}}.

\bibitem{Kachru:2003aw}
S.~Kachru, R.~Kallosh, A.~D. Linde, and S.~P. Trivedi, ``{De Sitter vacua in
  string theory},'' {\em Phys.Rev.} {\bf D68} (2003) 046005,
\href{http://arXiv.org/abs/hep-th/0301240}{{\tt hep-th/0301240}}.

\bibitem{MR522939}
F.~A. Bogomolov, ``Holomorphic tensors and vector bundles on projective
  manifolds,'' {\em Izv. Akad. Nauk SSSR Ser. Mat.} {\bf 42} (1978), no.~6,
  1227--1287, 1439.

\bibitem{Douglas:2006jp}
M.~R. Douglas, R.~Reinbacher, and S.-T. Yau, ``{Branes, bundles and attractors:
  Bogomolov and beyond},''
\href{http://arXiv.org/abs/math/0604597}{{\tt math/0604597}}.

\bibitem{Andreas:2010hv}
B.~Andreas and G.~Curio, ``{Spectral Bundles and the DRY-Conjecture},'' {\em
  J.Geom.Phys.} {\bf 62} (2012) 800--803,
\href{http://arXiv.org/abs/1012.3858}{{\tt 1012.3858}}.

\bibitem{Andreas:2011zs}
B.~Andreas and G.~Curio, ``{On the Existence of Stable bundles with prescribed
  Chern classes on Calabi-Yau threefolds},''
\href{http://arXiv.org/abs/1104.3435}{{\tt 1104.3435}}.

\bibitem{Blumenhagen:2005ga}
R.~Blumenhagen, G.~Honecker, and T.~Weigand, ``{Loop-corrected
  compactifications of the heterotic string with line bundles},'' {\em JHEP}
  {\bf 0506} (2005) 020,
\href{http://arXiv.org/abs/hep-th/0504232}{{\tt hep-th/0504232}}.

\bibitem{Blumenhagen:2005zg}
R.~Blumenhagen, G.~Honecker, and T.~Weigand, ``{Non-Abelian brane worlds: The
  Heterotic string story},'' {\em JHEP} {\bf 0510} (2005) 086,
\href{http://arXiv.org/abs/hep-th/0510049}{{\tt hep-th/0510049}}.

\bibitem{Blumenhagen:2006ux}
R.~Blumenhagen, S.~Moster, and T.~Weigand, ``{Heterotic GUT and standard model
  vacua from simply connected Calabi-Yau manifolds},'' {\em Nucl.Phys.} {\bf
  B751} (2006) 186--221,
\href{http://arXiv.org/abs/hep-th/0603015}{{\tt hep-th/0603015}}.

\bibitem{Blumenhagen:2006wj}
R.~Blumenhagen, S.~Moster, R.~Reinbacher, and T.~Weigand, ``{Massless Spectra
  of Three Generation U(N) Heterotic String Vacua},'' {\em JHEP} {\bf 0705}
  (2007) 041,
\href{http://arXiv.org/abs/hep-th/0612039}{{\tt hep-th/0612039}}.

\bibitem{Weigand:2006yj}
T.~Weigand, ``{Compactifications of the heterotic string with unitary
  bundles},'' {\em Fortsch.Phys.} {\bf 54} (2006)
963--1077.

\bibitem{Anderson:2009nt}
L.~B. Anderson, J.~Gray, A.~Lukas, and B.~Ovrut, ``{Stability Walls in
  Heterotic Theories},'' {\em JHEP} {\bf 0909} (2009) 026,
\href{http://arXiv.org/abs/0905.1748}{{\tt 0905.1748}}.

\bibitem{Anderson:2010tc}
L.~B. Anderson, J.~Gray, and B.~Ovrut, ``{Yukawa Textures From Heterotic
  Stability Walls},'' {\em JHEP} {\bf 1005} (2010) 086,
\href{http://arXiv.org/abs/1001.2317}{{\tt 1001.2317}}.

\bibitem{Anderson:2011ns}
L.~B. Anderson, J.~Gray, A.~Lukas, and E.~Palti, ``{Two Hundred Heterotic
  Standard Models on Smooth Calabi-Yau Threefolds},'' {\em Phys.Rev.} {\bf D84}
  (2011) 106005,
\href{http://arXiv.org/abs/1106.4804}{{\tt 1106.4804}}.

\bibitem{Anderson:2012yf}
L.~B. Anderson, J.~Gray, A.~Lukas, and E.~Palti, ``{Heterotic Line Bundle
  Standard Models},'' {\em JHEP} {\bf 1206} (2012) 113,
\href{http://arXiv.org/abs/1202.1757}{{\tt 1202.1757}}.

\bibitem{Anderson:2011vy}
L.~B. Anderson, J.~Gray, A.~Lukas, and E.~Palti, ``{Heterotic standard models
  from smooth Calabi-Yau three-folds},'' {\em PoS} {\bf CORFU2011} (2011)
096.

\bibitem{Donagi:1998xe}
R.~Donagi, A.~Lukas, B.~A. Ovrut, and D.~Waldram, ``{Nonperturbative vacua and
  particle physics in M theory},'' {\em JHEP} {\bf 9905} (1999) 018,
\href{http://arXiv.org/abs/hep-th/9811168}{{\tt hep-th/9811168}}.

\bibitem{Donagi:1999gc}
R.~Donagi, A.~Lukas, B.~A. Ovrut, and D.~Waldram, ``{Holomorphic vector bundles
  and nonperturbative vacua in M theory},'' {\em JHEP} {\bf 9906} (1999) 034,
\href{http://arXiv.org/abs/hep-th/9901009}{{\tt hep-th/9901009}}.

\bibitem{Donagi:1999ez}
R.~Donagi, B.~A. Ovrut, T.~Pantev, and D.~Waldram, ``{Standard models from
  heterotic M theory},'' {\em Adv.Theor.Math.Phys.} {\bf 5} (2002) 93--137,
\href{http://arXiv.org/abs/hep-th/9912208}{{\tt hep-th/9912208}}.

\bibitem{Donagi:2003tb}
R.~Donagi, B.~A. Ovrut, T.~Pantev, and R.~Reinbacher, ``{SU(4) instantons on
  Calabi-Yau threefolds with Z(2) x Z(2) fundamental group},'' {\em JHEP} {\bf
  0401} (2004) 022,
\href{http://arXiv.org/abs/hep-th/0307273}{{\tt hep-th/0307273}}.

\bibitem{Braun:2004xv}
V.~Braun, B.~A. Ovrut, T.~Pantev, and R.~Reinbacher, ``{Elliptic Calabi-Yau
  threefolds with Z(3) x Z(3) Wilson lines},'' {\em JHEP} {\bf 0412} (2004)
  062,
\href{http://arXiv.org/abs/hep-th/0410055}{{\tt hep-th/0410055}}.

\bibitem{Donagi:2004ub}
R.~Donagi, Y.-H. He, B.~A. Ovrut, and R.~Reinbacher, ``{The Spectra of
  heterotic standard model vacua},'' {\em JHEP} {\bf 0506} (2005) 070,
\href{http://arXiv.org/abs/hep-th/0411156}{{\tt hep-th/0411156}}.

\bibitem{MR923487}
C.~Schoen, ``On fiber products of rational elliptic surfaces with section,''
  {\em Math. Z.} {\bf 197} (1988), no.~2, 177--199.

\bibitem{Ovrut:2002jk}
B.~A. Ovrut, T.~Pantev, and R.~Reinbacher, ``{Torus fibered Calabi-Yau
  threefolds with nontrivial fundamental group},'' {\em JHEP} {\bf 0305} (2003)
  040,
\href{http://arXiv.org/abs/hep-th/0212221}{{\tt hep-th/0212221}}.

\bibitem{Braun:2007tp}
V.~Braun, M.~Kreuzer, B.~A. Ovrut, and E.~Scheidegger, ``Worldsheet Instantons,
  Torsion Curves, and Non-Perturbative Superpotentials,'' {\em Phys. Lett.}
  {\bf B649} (2007) 334--341,
\href{http://arXiv.org/abs/hep-th/0703134}{{\tt hep-th/0703134}}.

\bibitem{Braun:2007xh}
V.~Braun, M.~Kreuzer, B.~A. Ovrut, and E.~Scheidegger, ``Worldsheet instantons
  and torsion curves. Part A: Direct computation,'' {\em JHEP} {\bf 10} (2007)
  022,
\href{http://arXiv.org/abs/hep-th/0703182}{{\tt hep-th/0703182}}.

\bibitem{Braun:2007vy}
V.~Braun, M.~Kreuzer, B.~A. Ovrut, and E.~Scheidegger, ``Worldsheet Instantons
  and Torsion Curves, Part B: Mirror Symmetry,'' {\em JHEP} {\bf 10} (2007)
  023,
\href{http://arXiv.org/abs/arXiv:0704.0449 [hep-th]}{{\tt arXiv:0704.0449
  [hep-th]}}.

\bibitem{Gomez:2005ii}
T.~L. Gomez, S.~Lukic, and I.~Sols, ``{Constraining the Kahler moduli in the
  heterotic standard model},'' {\em Commun.Math.Phys.} {\bf 276} (2007) 1--21,
\href{http://arXiv.org/abs/hep-th/0512205}{{\tt hep-th/0512205}}.

\bibitem{Green:1987sp}
M.~B. Green, J.~H. Schwarz, and E.~Witten, ``Superstring Theory. Vol. 1:
  Introduction,''. Cambridge, Uk: Univ. Pr. ( 1987) 469 P. ( Cambridge
  Monographs On Mathematical Physics).

\bibitem{Green:1987mn}
M.~B. Green, J.~H. Schwarz, and E.~Witten, ``Superstring Theory. Vol. 2: Loop
  Amplitudes, Anomalies And Phenomenology,''. Cambridge, Uk: Univ. Pr. ( 1987)
  596 P. ( Cambridge Monographs On Mathematical Physics).

\bibitem{Blumenhagen:2005pm}
R.~Blumenhagen, G.~Honecker, and T.~Weigand, ``{Supersymmetric (non-)Abelian
  bundles in the Type I and SO(32) heterotic string},'' {\em JHEP} {\bf 0508}
  (2005) 009,
\href{http://arXiv.org/abs/hep-th/0507041}{{\tt hep-th/0507041}}.

\bibitem{Distler:2007av}
J.~Distler and E.~Sharpe, ``{Heterotic compactifications with principal bundles
  for general groups and general levels},'' {\em Adv.Theor.Math.Phys.} {\bf 14}
  (2010) 335--398,
\href{http://arXiv.org/abs/hep-th/0701244}{{\tt hep-th/0701244}}.

\bibitem{Distler:1987ee}
J.~Distler and B.~R. Greene, ``{Aspects of (2,0) String Compactifications},''
  {\em Nucl.Phys.} {\bf B304} (1988)
1.

\bibitem{Freed:1986zx}
D.~Freed, ``{Determinants, Torsion, and Strings},'' {\em Commun.Math.Phys.}
  {\bf 107} (1986)
483--513.

\bibitem{Witten:1985mj}
E.~Witten,
``{Global Anomalies in String Theory},''.

\end{thebibliography}\endgroup

\end{document}